\documentclass[twocolumn,superscriptaddress]{revtex4-1}
\pdfoutput=1 
\usepackage{graphicx}
\usepackage{amsmath}
\usepackage{amsfonts}
\usepackage{amssymb}

\begin{document}

\title{Sorting photon wave packets using temporal-mode interferometry
  based on multiple-stage quantum frequency conversion}

\author{D. V. Reddy}
\author{M. G. Raymer}
\email{raymer@uoregon.edu}
\affiliation{Oregon Center for Optics, Department of Physics, University of Oregon, Eugene, Oregon 97403, USA}
\author{C. J. McKinstrie}
\affiliation{Applied Communication Sciences, Red Bank, New Jersey 07701, USA}

\date{\today}

\begin{abstract}
All classical and quantum technologies that encode in and retrieve
information from optical fields rely on the ability to selectively
manipulate orthogonal field modes of light. Such manipulation can be
achieved with high selectivity for polarization modes and
transverse-spatial modes. For the time-frequency degree of freedom,
this could efficiently be achieved for a limited choice of
approximately orthogonal modes, i.e. non-overlapping bins in time or
frequency. We recently proposed a method that surmounts the
selectivity barrier for sorting arbitrary orthogonal temporal modes
[Opt. Lett. {\bf 39}, 2924 (2014)] using cascaded interferometric
quantum frequency conversion in nonlinear optical media. We call this
method temporal-mode interferometry, as it has a close resemblance to
the well-known separated-fields atomic interferometry method
introduced by Ramsey. The method has important implications for
quantum memories, quantum dense coding, quantum teleportation, and
quantum key distribution. Here we explore the inner workings of the
method in detail, and extend it to multiple stages with a concurrent
asymptotic convergence of temporal-mode selectivity to unity. We also
complete our analysis of pump-chirp compensation to counter
pump-induced nonlinear phase-modulation in four-wave mixing
implementations.
\end{abstract}


\maketitle

\section{I. Introduction}

The state of a photon is fully characterized by its helicity, and the
three components of its momentum. For guided photons or photons in
beam-like geometries, the same degrees of freedom may be stated as
polarization, transverse-mode profile, and energy (or frequency)
\cite{smi07}. Polarization spans a two-dimensional state space,
allowing definitions of two-tuple orthogonal basis sets, for example
perpendicular linear polarization states. Sorting and the application
of unitary transformations on any choice of bases is accomplished
through combinations of existing tools such as polarizing beam
splitters, half-wave plates, and quarter-wave plates. Transverse modes
are two-dimensional functions, and may be expressed in the basis of
orbital angular momentum, transverse-spatial parity, or integer
indices denoting eigenfunctions of some guiding geometry. In contrast
to polarization, transverse modes offer a multidimensional state
space, and their selective manipulation has been an active subject of
recent studies
\cite{sas03,abou07,winzer09,bar10,pad10,yao11,wang12,wil13}.

For traveling (longitudinally unconstrained) fields, energy (or
frequency) is a continuous-variable degree of freedom. This spans an
infinite dimensional space, and high-fidelity selective sorting (and
by extension, any unitary operation) on arbitrary states in this space
has traditionally been limited to specific choices of basis sets.  If
the considered states have disjoint spectra, this is accomplished
using a prism or diffraction grating. That is the basis of
wavelength-division multiplexing in optical communication
systems. Sorting has also been accomplished for spread-spectrum
distributions of orthogonal frequency subcarriers via
orthognal-frequency-division-multiplexing (OFDM) \cite{yang11}, and
for partially overlapping temporal pulses of a specific shape and
temporal offset via orthogonal-time-division-multiplexing (OTDM)
\cite{nakazawa14}. But orthogonal states can also be defined as
superpositions over a continuum of frequencies with overlapping
spectra. By Fourier transformation, such states can be represented in
the time domain, thereby introducing the idea of `temporal modes'--a
set of orthogonal functions of time that form a complete basis for
representing an arbitrary state in the energy degree of freedom
\cite{tit66,smi07}.

Temporal modes (TMs) are a prime contender for the preferred basis for
mapping qubits or higher-dimensional qudits onto photons, because
their orthogonality is immune to chromatic dispersion and polarization
rotation. TMs are ideal for transmission in waveguides and optical
fibers with no cross-talk: since all TMs at a given central frequency
suffer the same unitary evolution under linear dispersion, their
orthogonality is maintained in long-distance quantum communication
through time-stationary guided media.

The most promising approach to the sorting of light into a discrete
set of TMs has been nonlinear wave mixing with pulsed pumps
\cite{mcg10a,eck11}. From a general mathematical standpoint analogous
to that in \cite{wal96}, it is clear that temporal wave-packet
multiplexing is possible only by using a temporally nonstationary
process that acts on field amplitudes, not intensities. This is
inherently true in strong-pump driven nonlinear wave mixing, where the
weak signal fields need to be synchronous with a device clock, defined
by the arrival time of the pump pulse. Nonzero relative group
velocities between participating fields also allow the device to
``sample'' or ``measure'' the complex temporal amplitude of the
signals via a continously shifting inter-pulse temporal overlap, which
is essential if it needs to descriminate between orthogonal modes. The
complex envelopes of the strong pump pulses provide for a means of
programming the device to sort user-defined temporal modes. Such a
system can be called a complex-pulse-envelope gate, or pulse gate for
short \cite{eck11}. It has also been called a field-orthogonal TM
sorter \cite{red14}.

It was first pointed out in \cite{mcg10a} that quantum frequency
conversion (QFC) by four-wave mixing (FWM) is partially TM-selective
when using short pump pulses. The same is true also for QFC by
three-wave mixing (TWM) driven by a single pump pulse, as pointed out
by Eckstein et al, and proposed as a general method of TM sorting and
multiplexing for quantum information applications \cite{eck11}. The
idea is to design the process so that only a single temporal mode
among the orthogonal set has its frequency converted; then a simple
frequency-selective component (diffraction grating, prism, etc.) could
easily separate the converted wave packet from the unconverted
ones. Recent studies furthered the concept of using optical frequency
conversion (FC) by nonlinear wave mixing for creating
TM-mode-selective devices \cite{col12,mej12,red13,huang14}. The role
of FC in quantum optical communication networks has been reviewed in
\cite{ptd12}.

Ideally, the device should frequency convert a user-specified temporal
mode component from the input field into the output mode in a
different frequency band without `contamination' by other temporal
mode components of the input field in the orthogonal subspace. But it
was found through careful modeling that the basic single-stage QFC
process is fundamentally limited to a TM `selectivity' of no greater
than about $0.8$, strongly limiting its usefulness
\cite{eck11,col12,red13,silb14,huang14}. High selectivity is a
necessary condition for scalability in quantum applications, and a
value of $0.8$ falls short of error-tolerance estimates in both
quantum communication and computing. The limit exists as a fundamental
feature of group-velocity mismatched inter-pulse dynamics in a
dispersive medium, as will be shown. We discovered that a
generalization of the QFC schemes (either FWM or TWM) can circumvent
this limit and achieve essentially perfect TM sorting, with a
selectivity approaching unity \cite{red14}. The new scheme uses two or
more spatially separated stages of QFC, with the effects of the stages
adding coherently. We named this scheme {\it temporal-mode
  interferometry} (TMI), as it has a close resemblance to the
separated-fields atomic interferometry method introduced by Ramsey
\cite{ramsey50}. In that case, an atom passes through two separated
laser beams. The first field partially excites the atom (for example,
to $50$\% probability), and the second field, depending on its phase
relative to that of the oscillating atomic dipole resulting from the
first field, either further excites the atom (for example, to $100$\%)
or fully deexcites the atom back to its ground state. The present case
of TMI is more complex, in that the traveling pulse is not localized
at a point like a typical atom, but undergoes temporal shaping as its
frequency is being converted. Time-domain versions of Ramsey
interference are also known in the contexts of photon-echo experiments
and atomic-fountain clocks \cite{mos79,chu89}.

The dependence of conversion efficiency on the relative phases of the
signal and control fields is what earns the TMI method the name
interferometry. It is useful for real-time control and switching of
various temporal modes. The concept of two-stage QFC, and its analogy
to Ramsey interferometry, was subsequently independently developed by
Clemmen et al \cite{gaeta14}, but only in the context of
continuous-wave fields, so in this case it lacks the TM selectivity
that is the main thrust of the present discussion.

This paper details and expands the work reported in \cite{red14}. In
particular, we present the theory of inter-stage mode matching. We
explore the dependence of mode selectivity on the ratio of
interaction-time and pump temporal duration in TWM. We provide the
theory of pump-chirp compensation in FWM, including a family of
analytical expressions for pump chirps. Finally, we give numerical
demonstrations of multistage (more than two) TMI and the associated
asymptotic gains in mode selectivity.

\section{II. Temporal modes}

Temporal modes are a set of complex functions that are orthogonal in
time (or frequency), which form a complete basis for representing an
arbitrary state in the energy degree of freedom
\cite{smi07,tit66}. The state can be expressed as a coherent
superposition of many possible frequency states,

\begin{align}
  &|\psi_j\rangle = \frac{1}{2\pi}\int d\omega \widetilde{\psi}_j(\omega)\widehat{a}^\dagger(\omega)|vac\rangle = \widehat{b}_j^\dagger|vac\rangle,\\
  &[\widehat{a}(\omega),\widehat{a}^\dagger(\omega')]=2\pi\delta(\omega-\omega'),\quad [\widehat{b}_j,\widehat{b}^\dagger_k]=\delta_{j,k},\notag
\end{align}

\noindent where $\widehat{a}^\dagger(\omega)$ creates a monochromatic
photon with frequency $\omega$. By Fourier transform, the same state
can be expressed as a coherent quantum superposition state of many
possible `creation times,' that is

\begin{equation}
  |\psi_j\rangle=\int dt\psi_j(t)\widehat{A}^\dagger(t)|vac\rangle,
\end{equation}

\noindent where $\widehat{A}^\dagger(t)$ creates a broadband photon at
time $t$. The TM mode functions $\psi_j(t)$ fully overlap in time, but
are orthogonal with respect to a time integral. Likewise in the
frequency domain. The mode basis sets $\{\psi_j(t)\}$ (or
$\{\widetilde{\psi}_j(\omega)\}$) can be any user-defined orthonormal
set. The integer index is in principle unbound, reflecting the
infinite dimensionality of a continuous variable degree of
freedom. For exceptionally broadband TMs, covering essentially the
entire visible spectrum, exact orthogonality is not ensured
\cite{smi07}. But this subtlety will have negligible effects on
practical systems where channels are defined with bandwidths that are
narrow compared to the carrier frequencies.

\begin{figure}[htb]
\centering
\includegraphics[width=\columnwidth]{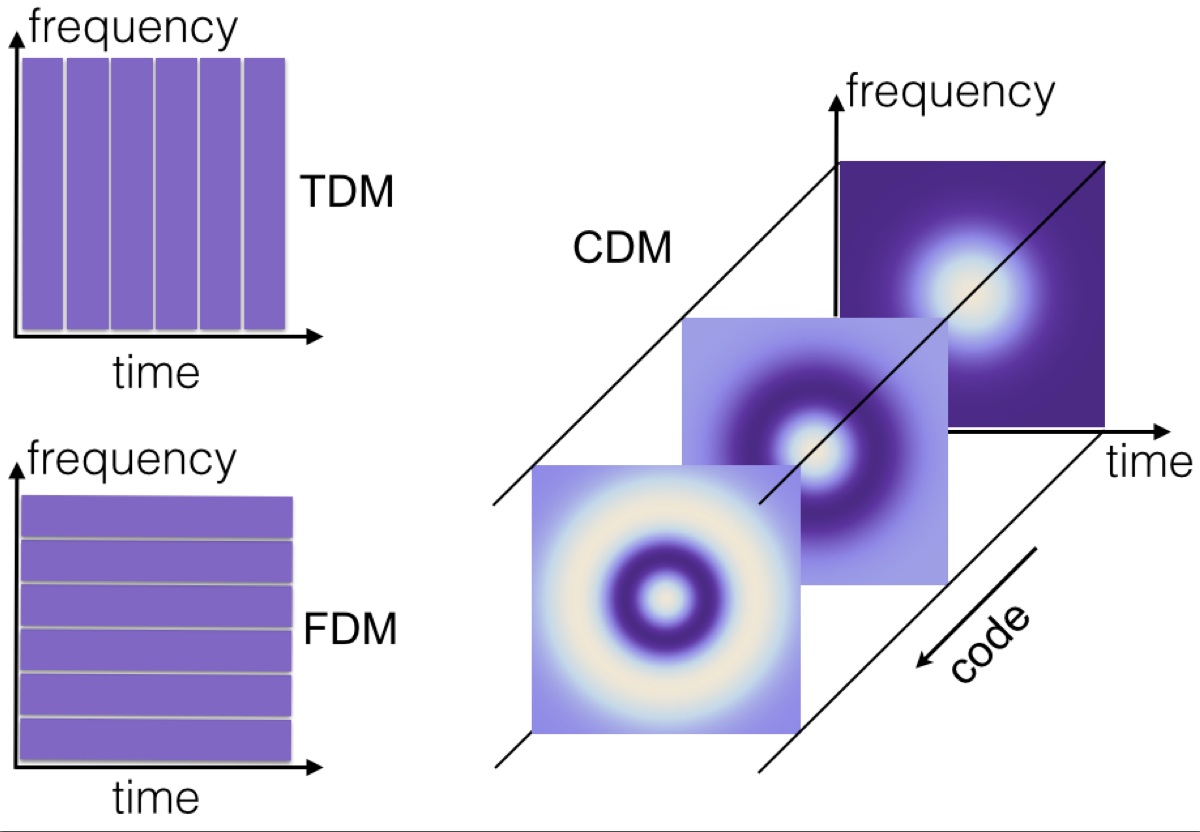}
\caption{Time- and frequency-division multiplexing being contrasted
  with code-division multiplexing (CDM), with different codes
  represented by time-frequency Wigner functions for orthogonal
  temporal modes. The choice of the temporal-mode set is not unique.}
\label{fig01}
\end{figure}

TM-based multiplexing has more in common with code-division
multiplexing, as the different overlapping functions serve as
spread-spectrum codes that mark information channels. This is in
contrast with conventional time- or frequency-based optical
multiplexing, which use either separated short pulses or narrow
spectral windows to define different information channels
(fig. \ref{fig01}).

\section{III. Single-stage quantum frequency conversion}

Here, we present the theory for QFC via three-wave mixing (TWM)
\cite{huang92,vand04,albota04,rous04,rakher10} and four-wave mixing
(FWM) \cite{mcg10a,cla13}. To model the process, we designate the
participating frequency channels by {\it s} and {\it r} for the signal
photons, and {\it p} (and {\it q} ) for the strong pump field(s). We
denote square-normalized classical field envelopes in the {\it
  j} channel by $A_j(z,t)$. Quantum mechanically, $A_j(z,t)$ represent
the field annihilation operators for a wave-packet mode of that
central frequency ($\omega_j$) and envelope shape \cite{smi07} and
obey the same equations of motion as their classical counterparts
\cite{mcg10a,red13}. Then the equations of motion for QFC are
\cite{myers95,mej12b}:

\begin{subequations}
\label{eomfwm}
\begin{align}
(\partial_z+\beta'_p\partial_t)A_p&=i(\gamma/2)\delta_F\left[\left|A_p\right|^2+2\left|A_q\right|^2\right]A_p,\\
(\partial_z+\beta'_q\partial_t)A_q&=i(\gamma/2)\delta_F\left[2\left|A_p\right|^2+\left|A_q\right|^2\right]A_q,\\
(\partial_z+\beta'_r\partial_t)A_r&=i\gamma A_pA_q^*A_s\notag\\
&+i\gamma\delta_F\left[\left|A_p\right|^2+\left|A_q\right|^2\right]A_r,\\
(\partial_z+\beta'_s\partial_t)A_s&=i\gamma A_p^*A_qA_r\notag\\
&+i\gamma\delta_F\left[\left|A_p\right|^2+\left|A_q\right|^2\right]A_s,
\end{align}
\end{subequations}

\noindent where $\beta'_j\equiv[d\beta/d\omega]|_{\omega=\omega_j}$
are the group slownesses (inverse group velocities), and we have
neglected higher-order dispersion, which is valid for sufficiently
narrow-band pulses. The self- and cross-phase modulation factor
$\delta_F$ is $0$ for TWM and $1$ for FWM. $\gamma$ is the product of
the effective $\chi^{(2)}$ or $\chi^{(3)}$ nonlinearity and the square
roots of the pump-pulse energies. For TWM, $A_q(z,t)\equiv 1$. The
presence or absence of complex-conjugation of pump amplitudes follows
the convention that $\omega_r = \omega_p + \omega_s - \omega_q$ for
FWM, and $\omega_r = \omega_p + \omega_s$ for TWM. The solution for
Eq. (\ref{eomfwm}) can be written in terms of Green functions
$G_{ij}(t,t')$. For $j\in \{r,s\}$ and medium length $l$:

\begin{equation}
A_j(l,t)=\int^\infty_{-\infty}\mathrm{d}t'\sum\limits_{k=r,s}G_{jk}(t,t')A_k(0,t').\label{eq07}
\end{equation}

The four kernels \cite{col12} $G_{ij}(t,t')$ comprise a generalized
beam-splitter transformation, representing frequency conversion or
non-conversion of each of the two possible input fields of distinct
carrier frequencies \cite{ray10}. The functional forms of the Green
function kernels are dependent on the shapes of the {\it p} (and {\it
  q}) pump pulse(s). To determine the natural temporal modes for the
problem, we numerically perform a singular-value decomposition (SVD),
also called a Schmidt decomposition, for each $G_{ij}(t,t')$. This
yields a set of singular-value functions (temporal Schmidt modes) with
associated singular values (Schmidt coefficients), which we denote by
$\rho_n$ and $\tau_n$, which satisfy
$|\rho_n|^2+|\tau_n|^2=1$. $|\rho_n|^2$ is the frequency conversion
efficiency (CE) of the $n$th mode, and can be interpreted as the
probability of QFC in the case of single photons. The modes are
numbered in decreasing order of the Schmidt coefficients; the mode
with the largest Schmidt coefficient ({\it i.e.} the first Schmidt
mode, $n=1$) is the target TM (expected to be selectively
converted). We denote by $\phi_n(t)$ the {\it s}-input modes, and
$\Psi_n(t)$ are {\it r}-output modes. In addition, there are {\it
  r}-input modes $\psi_n(t)$ and {\it s}-output modes $\Phi_n(t)$.

All the Green function (GF) kernels can be expressed in terms of these
four mode sets as follows \cite{ray10}:

\begin{subequations}
\begin{align}
G_{rr}(t,t')&=\sum\limits^\infty_{n=1}\tau_n\Psi_n(t)\psi_n^*(t'),\\
G_{rs}(t,t')&=\sum\limits^\infty_{n=1}\rho_n\Psi_n(t)\phi_n^*(t'),\\
G_{sr}(t,t')&=-\sum\limits^\infty_{n=1}\rho_n\Phi_n(t)\psi_n^*(t'),\\
G_{ss}(t,t')&=\sum\limits^\infty_{n=1}\tau_n\Phi_n(t)\phi_n^*(t').
\end{align}\label{svdeq}
\end{subequations}

As mentioned earlier, the individual kernals are not unitary, but
taken together, the complete GF is. The objective is to design a QFC
device that can selectively frequency convert or `drop' the first
Schmidt mode (target TM) with unit efficiency, whilst allowing $100$\%
unconverted transmission of all orthogonal modes. The principal figure
of merit for such a drop device is the `selectivity' \cite{red13},
defined as:

\begin{equation}
  S = \frac{|\rho_1|^4}{\sum\limits^\infty_{j=1}|\rho_j|^2}.
\end{equation}

\begin{figure}[htb]
\centering
\includegraphics[width=\columnwidth]{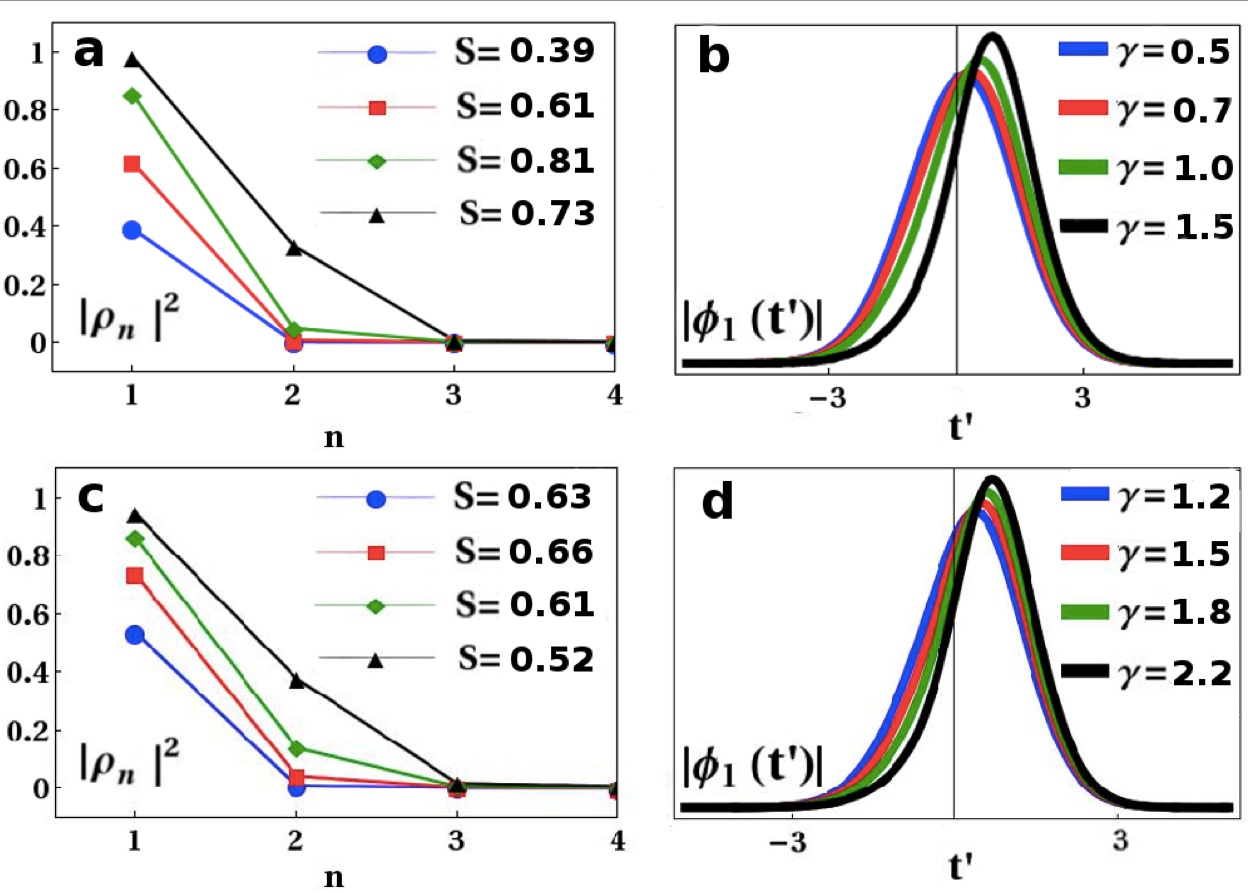}
\caption{ Single-stage TWM (a,b) and FWM (c,d) dominant
  Schmidt-mode conversion efficiencies $|\rho_n|^2$ (a,c) for various
  pump powers, with selectivities $S$ listed in the legend. (b,d)
  s-channel input Schmidt modes, illustrating increased temporal
  skewness with increasing $\gamma$. The values of $t'$ are relative to
  a $|\beta'_r-\beta'_s|l$ of (b) $20$ and (d) $10$, where $l$ is the
  medium length. Pump-{\it p} is Gaussian in shape.}
\label{fig02}
\end{figure}

The best previous attempts at optimizing selectivity using a
single-stage QFC scheme converged on frequency choices that are
group-velocity matched, namely
$\beta'_p=\beta'_s\neq\beta'_r$($=\beta'_q$)\cite{mej12b,col12,red13}. For
FWM, this condition is easily accessed by situating the channel
frequencies on either side of the zero-dispersion wavelength of the
fiber, which automatically grants phase matching for a tunable set of
frequencies \cite{mcg10a}. In addition, for optimum selectivity
results in FWM, the medium is required to be long enough for a
complete inter-pump-pulse collision (no overlap to
no overlap) \cite{mej12b}. However, both TWM and FWM schemes have
encountered limits that are universal to generic systems governed by
coupled-mode equations such as Eq. (\ref{eomfwm}), and have yielded
selectivities limited to around $0.8$, an example of which is shown in
Fig. \ref{fig02}. The single-stage selectivity limit arises from in
the interaction between pulses convecting through each other at
different speeds. The nonlinear wave-mixing is perfectly phase matched
only at the central channel frequencies, and the phase mismatch
acquires opposite relative signs for spectral components of the pulses
below and above the central frequencies. This causes the QFC
contributions of the two halves of the spectrum to beat relative to
each other, which manifest as non-separable time-domain oscillations
in the GF kernels \cite{red13} similar to Burnham-Chiao ringing, which
occurs when a weak, resonant optical pulse propagates through an
atomic vapor \cite{Burnham1969}. At high conversion, the Schmidt-mode
profiles become temporally skewed relative to the
group-velocity-matched pump shapes. These distortions are minimal at
lower CE, enabling our two-stage scheme.

\begin{figure*}[htb]
\centering
\includegraphics[width=\linewidth]{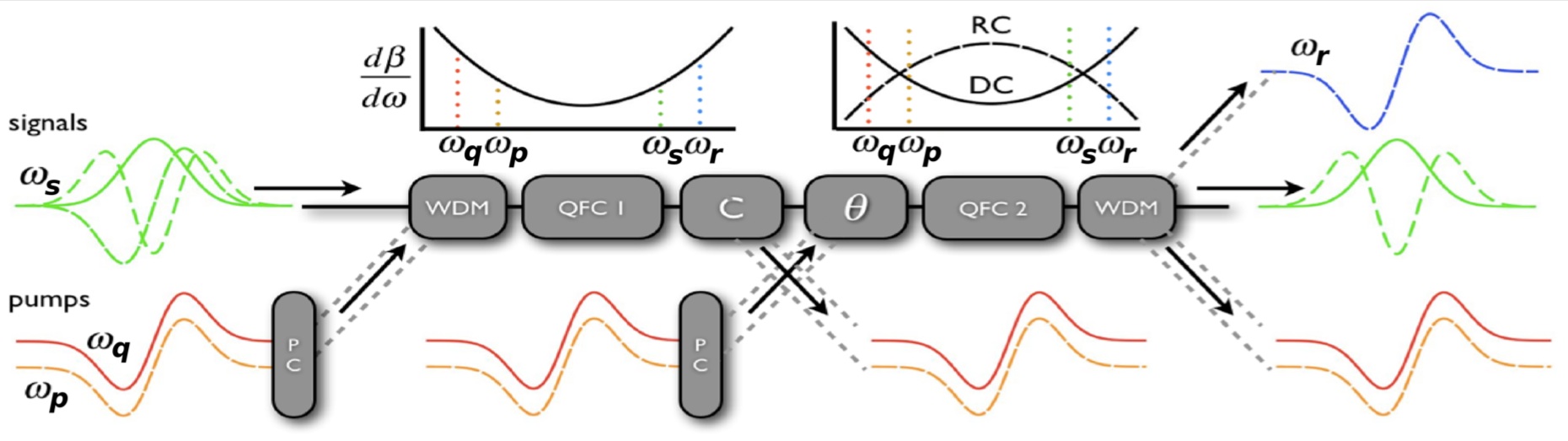}
\caption{ Temporal-mode interferometer using two
  nonlinear media (QFC 1 and QFC 2) with identical (DC) or
  opposite-sign (RC) dispersion. Appropriate choices for pump-pulse
  shapes, pump powers and the phase shift $\theta$ will selectively
  frequency convert a specific {\it s} (green, $\omega_s$) TM into an
  {\it r} (blue, $\omega_r$) TM at a different central frequency,
  while not affecting temporally-orthogonal {\it s}-input TMs. The
  pump {\it q} is used only for $\chi^{(3)}$ implementations. WDM
  stands for wavelength-division multiplexer. PC stands for pre-chirp
  modules, which are necessary for $\chi^{(3)}$ implementations. The
  coupler C contains frequency dependent delays for the DC
  case. Reproduced from \cite{red14}, with permission.}
\label{fig03}
\includegraphics[width=\linewidth]{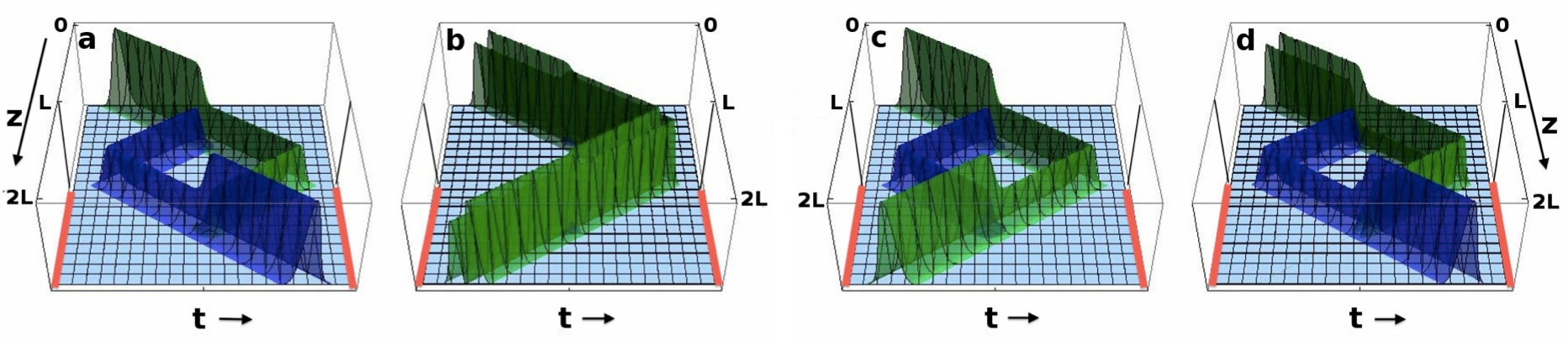}
\caption{ Temporal-mode interferometry with the two
  interferometer `arms' being the frequency channels {\it s} (green,
  lighter shade) and {\it r} (blue, darker shade), undergoing two
  complete collisions in the two fibers and exchanging energy. This
  visualization plots the color-coded signal-field intensities in the
  average velocity frame with populated green input and empty blue
  input at $z = 0$, for the RC configuration found by numerical
  solution of the equations of motion. (a) Both pumps are Gaussian,
  and the first Schmidt-mode green-to-blue conversion is nearly 100\%
  at phase-shift $\theta = 0$. (b) Both pumps are Gaussian, and the
  second Schmidt-mode conversion is nearly 0\%. (c) Both pumps are
  Gaussian, and the first Schmidt-mode conversion efficiency is
  suppressed to zero at phase-shift $\theta = \pi$. (d) Pump {\it q}
  is Gaussian, and the shape of pump {\it p} is tailored to convert
  the green input mode from (b) into a Gaussian-like blue output with
  nearly 100\% efficiency at $\theta = 0$.}
\label{fig04}
\end{figure*}

\section{IV. Temporal-mode interferometry}

Our scheme, which we call temporal-mode interferometry, uses cascaded
stages of QFC, as illustrated in Fig. \ref{fig03}. The method provides
close to $100$\% TM selectivity, opening the door for manipulation of
TM qubits or multi-level qudits, and may also be used for TM channel
multiplexing in classical optical telecommunications. In the figure we
color code the participating frequency channels {\it s} in green, and
{\it r} in blue for the signal photons, as well as {\it p} (and {\it
  q} ) the strong pump field(s) in orange (and red). Our method
exploits the fact that using single-stage QFC, it is possible to
discriminate orthogonal TMs nearly perfectly up to a conversion
efficiency (CE) of about 50\% \cite{col12,red13}. This motivates a
two-stage interferometric scheme, in which each stage is configured
for $50$\% CE for the target TM, and functions as a $50/50$ beam
splitter with the {\it r}- and {\it s}-frequency channels representing
its two input and output arms. An {\it s}-input photon in the target
TM will be $50$\% frequency converted in the first stage into the {\it
  r} channel with a phase picked up from the pump field(s). If all
fields are allowed to participate in QFC in the second stage with the
right relative phases, the two effective beam splitters will function
as a frequency-shifting Mach-Zehnder (or Ramsey) interferometer,
allowing for complete forward or backward conversion of the state of
the photon, depending on the value of the additional relative phase
$\theta$ introduced between the stages. Between the two stages, the
target TM component of the signal photon is in a multicolor
superposition of the signal-idler frequency bands ({\it i.e.} a color
qubit)\cite{pfister04,col08}. This interferometric scheme was first
published by us in a previous work \cite{red14}. Phase coherence
between frequency-converted and unconverted components of light has
since been experimentally demonstrated for weak-coherent states by
Clemmen et al \cite{gaeta14}, who independently arrived at the
frequency-conversion interferometry idea. However, their
implementation used continuous pumps and so is not TM selective. In
our case, the interferometric frequency-conversion effect operates
only on the target TM, as the orthogonal modes have negligible CE in
both stages.

There are two configurations of interest: 1) `reversed collision'
(RC), in which the dispersion in the second-stage is inverted relative
to that in the first stage, such that the relative group velocities of
the pulses are reversed, and 2) `double collision' (DC), in which the
dispersion in the second-stage is identical to that in the first. In
the DC case, the fast pulses must be time delayed relative to the slow
pulses in between the stages so that they walk through each other
again in the second stage. The inset in Fig. \ref{fig03} shows the
inverse group velocity vs. frequency for the two
configurations. Figure \ref{fig04} shows an example of TMI for the RC
configuration. The necessary time delays for the DC case are
implemented in the `coupler' labeled C in the figure. After the target
TM is frequency converted, it would be separated from the main beam
using a standard wavelength-division multiplexer (WDM). As shown
below, for TWM we predict in the RC case a TM selectivity of $0.9846$,
and in the DC case the selectivity is $0.9805$, far higher than can be
achieved with single-stage systems. High TM selectivity can also be
achieved using FWM, but will require chirp pre-compensation of the
pump pulses to be implemented in modules labeled PC, as described
below.

\subsection{Two-stage theory}

For the two-stage temporal-mode interferometry, the combined Green function
kernel $G_{rs}(t,t')$ is given by the interferometric equation \cite{red13}:

\begin{equation}
\begin{split}
G_{rs}(t,t')=\int_{-\infty}^\infty\mathrm{d}t''\big[G^{(2)}_{rs}(t,t'')G^{(1)}_{ss}(t'',t')\\
+e^{i\theta}G^{(2)}_{rr}(t,t'')G^{(1)}_{rs}(t'',t')\big],
\end{split}\label{eq15}
\end{equation}

\noindent where the superscripts indicate the FC process in stages 1
and 2. Functions and parameters lacking a superscript stage number
characterize the combined two-stage process as a whole. For each stage
separately, the `transmission' (no frequency change) coefficient
$\tau^{(\xi)}_n$ and the `reflection' (frequency change) coefficient
$\rho^{(\xi)}_n$, both taken to be real without loss of generality,
independently obey relations analogous to beam-splitter relations,
$\tau_n^{(\xi)2}+\rho^{(\xi)2}_n=1$. The stage-labeled input modes are
represented by lower-case functions $(\psi^{(\xi)},\phi^{(\xi)})$ and
output modes by upper case $(\Psi^{(\xi)},\Phi^{(\xi)})$.

For TWM and FWM (with complete pump collisions), described by
Eq. (\ref{eomfwm}), the Schmidt coefficients are independent of
pump-pulse shapes, and are determined by the value of $\gamma$. In
contrast, the Schmidt mode shapes are determined by the pump-pulse
shapes and the value of $\gamma$ \cite{col12}. For FWM, this allows
one to change the output signal shape to that of any temporal mode, if
desired. For TWM, which uses of a single pump, one can influence only
the Schmidt mode shapes of the group-velocity-matched signal channel.

The unitarity of the single-stage transformation imposes a pairing
between the Schmidt modes across the {\it r} and {\it s} channels for
that stage \cite{ray10}. If the input fields for a given stage
are represented as

\begin{subequations}
\begin{align}
&A_r^{(\xi)}(t')\big|_{\text{in}}=\sum\limits_na_n\psi^{(\xi)}_n(t'),\\
&A_s^{(\xi)}(t')\big|_{\text{in}}=\sum\limits_nb_n\phi^{(\xi)}_n(t'),
\end{align}
\end{subequations}

\noindent then the output fields are expressed as

\begin{subequations}
\begin{align}
&A_r^{(\xi)}(t)\big|_{\text{out}}=\sum\limits_n(\tau_n^{(\xi)}a_n+\rho^{(\xi)}_nb_n)\Psi^{(\xi)}_n(t),\\
&A_s^{(\xi)}(t)\big|_{\text{out}}=\sum\limits_n(\tau_n^{(\xi)}b_n-\rho^{(\xi)}_na_n)\Phi^{(\xi)}_n(t).
\end{align}
\end{subequations}

The expressions in brackets are equivalent to a beam-splitter
transformation, explaining why the QFC process is considered
background-free in principle \cite{ray10}. 

The operating principle of the TMI can be summarized simply as
follows. Consider the case that the input field to stage 1 is a single
temporal mode $A_s^{(1)}(t)|_{\text{in}}=b_n\phi_n^{(1)}(t)$, and the
{\it r}-input field is empty. Then, from Eqs. (\ref{eq07}) and
(\ref{svdeq}), the output fields of stage 1, and thus the input fields
of stage 2, are

\begin{subequations}
\begin{align}
&A_s^{(1)}(t)\big|_{\text{out}}=\tau^{(1)}_nb_n\Phi^{(1)}_n(t)=A_s^{(2)}(t)\big|_{\text{in}},\\
&A_r^{(1)}(t)\big|_{\text{out}}=\rho^{(1)}_nb_n\Psi^{(1)}_n(t)=e^{-i\theta}A_r^{(2)}(t)\big|_{\text{in}},
\end{align}
\end{subequations}

\noindent where a phase shift $\theta$ of the {\it r} field has been
introduced intentionally by the experimenter before the fields enter
stage 2. Then the output of stage 2 will be

\begin{subequations}\begin{align}
\begin{split}
A_s^{(2)}(t)\big|_{\text{out}}=&b_n\sum\limits_{m=1}^\infty\big(\tau^{(2)}_m\tau^{(1)}_n\mu_{m,n}\\
&-e^{i\theta}\rho^{(2)}_m\rho^{(1)}_n\eta_{m,n}\big)\Phi_m^{(2)}(t),\end{split}
\\
\begin{split}
A_r^{(2)}(t)\big|_{\text{out}}=&b_n\sum\limits_{m=1}^\infty\big(\rho^{(2)}_m\tau^{(1)}_n\mu_{m,n}\\
&+e^{i\theta}\tau^{(2)}_m\rho^{(1)}_n\eta_{m,n}\big)\Psi_m^{(2)}(t),
\end{split}\label{eq12b}
\end{align}\end{subequations}

\noindent where the `inter-stage mode overlaps' are defined as

\begin{subequations}
\begin{align}
&\mu_{m,n}=\int dt\phi_m^{(2)*}(t)\Phi_n^{(1)}(t),\label{eq13a}\\
&\eta_{m,n}=\int dt\psi_m^{(2)*}(t)\Psi^{(1)}_n(t).\label{eq13b}
\end{align}
\end{subequations}

If $|\mu_{n,n}|=|\eta_{n,n}|=1$, then we say the
processes in the two stages are temporally mode matched. This occurs
only if the dominant output modes of stage 1 coincide with the
corresponding input modes of stage 2 in each frequency channel, that
is $\Phi^{(1)}_n(t)=\phi_n^{(2)}(t)$, and
$\Psi^{(1)}_n(t)=\psi_n^{(2)}(t)$.

For the target mode $n = 1$, we wish to have
$A_s^{(2)}(t)\big|_{\text{out}}=0$ and
$A^{(2)}_r(t)\big|_{\text{out}}=b_1\Psi_1^{(2)}(t)$. This can only
occur if several conditions are met: {\bf 1)} the non-dominant Schmidt
coefficients $\rho^{(\xi)}_{n\ne 1}$ are nearly zero; {\bf 2)} the
dominant processes in the two stages are temporally mode matched; and
{\bf 3)} the dominant Schmidt coefficients are
$\rho_1^{(\xi)}=\tau_1^{(\xi)}=\sqrt{1/2}$. Then the phase $\theta$
needs to be adjusted to zero (or some other value if the mode overlaps
are complex). This gives
$(\tau^{(2)}_1\tau^{(1)}_1\mu_{1,1}-e^{i\theta}\rho^{(2)}_1\rho^{(1)}_1\eta_{1,1})=0$
and
$(\rho^{(2)}_1\tau^{(1)}_1\mu_{1,1}+e^{i\theta}\tau^{(2)}_1\rho^{(1)}_1\eta_{1,1})=1$. If
these conditions are met, then by varying the phase of the {\it r}
signal field between the fibers, either frequency channel in mode $n=1$ can be 100\% populated at the output of fiber 2.

As shown in the Numerical results section, the remaining (non-target)
modes ($n\neq 1$) have $\rho^{(\xi)}_n\approx 0,\tau^{(\xi)}_n\approx
1$. So, light in any one of these modes is not significantly frequency
converted in either stage, although there would be a change of
temporal mode shape if $\Phi^{(2)}_n(t)\neq\phi^{(1)}_n(t)$, creating
temporal mode distortion.

For many applications, especially those involving cascaded TM
operations, it would be highly beneficial to implement TMI without
temporal mode distortion of the non-converted signals. This is
achieved if the output Schmidt modes of stage 1 match the input
Schmidt modes of stage 2, that is $\Phi^{(1)}_n(t)=\phi^{(2)}_n(t)$,
and $\Psi^{(1)}_n(t)=\psi^{(2)}_n(t)$ for every $n$. The RC
configuration satisfies this requirement along with the conditions
$\Phi^{(2)}_n(t)=\phi^{(1)}_n(t)=\Phi_n(t)=\phi_n(t)$ and
$\Psi^{(2)}_n(t)=\psi^{(1)}_n(t)=\Psi_n(t)=\psi_n(t)$ ({\it i.e.} the
output Schmidt modes are identical to the input Schmidt modes for both
signal channels), leading to the elimination of temporal distortion if
the non-converted input modes are identical to the input Schmidt modes
of the system.

\subsection{Numerical results}

In order to derive the Green functions for QFC numerically, we
implemented a coupled-mode-equation solver that accepts arbitrary
input functions ($A_r(0,t')$, $A_s(0,t')$) as arguments, and computes
the resultant output functions ($A_r(l,t)$, $A_s(l,t)$) for
Eq. \ref{eomfwm}. This is achieved using a fourth-order Runge-Kutta
method. The solver iterates over differential steps in
pulse-propagation ($\Delta z$) from $z = 0$ to $l$ (medium
length). Every iteration consists of an upwinded $z$-propagation
scheme for all four pulses by a step ($\Delta z$), followed by a
Runge-Kutta implementation of the coupled nonlinear interaction, all
in spacetime domain.  We compute the GF by computing the outputs for
an orthogonal set of input `test signals'. To ellaborate, consider the
GF kernel and its Schmidt decomposition:
\begin{equation}
G_{rs}(t,t') = \sum\limits_j\rho_j\Psi_{j}(t)\phi^*_{j}(t').
\end{equation}
The objective is to calculate all the individual components ($\rho_j$,
$\Psi_{j}(t)$ , $\phi_{n}(t')$) on the right-hand side. We first pick
two arbitrary spanning sets of basis functions $\{B_{r,k}\}$ and
$\{\tilde{B}_{s,l}\}$ and re-express the Schmidt modes:
\begin{align}
\Psi_{j}(t)&=\sum\limits_kU_{jk}B_{r,k}(t);\quad\phi^*_{j}(t')=\sum\limits_lV_{jl}\tilde{B}^*_{s,l}(t'),\\
G_{rs}(t,t')&=\sum\limits_{k,l}\left[\sum\limits_jU_{jk}\rho_jV_{jl}\right]B_{r,k}(t)\tilde{B}^*_{s,l}(t')\notag\\
&=\sum\limits_{k,l}\left[\overline{G}_{rs}\right]_{kl}B_{r,k}(t)\tilde{B}^*_{s,l}(t').
\end{align}

\begin{figure*}[bht]
\centering
\includegraphics[width=\linewidth]{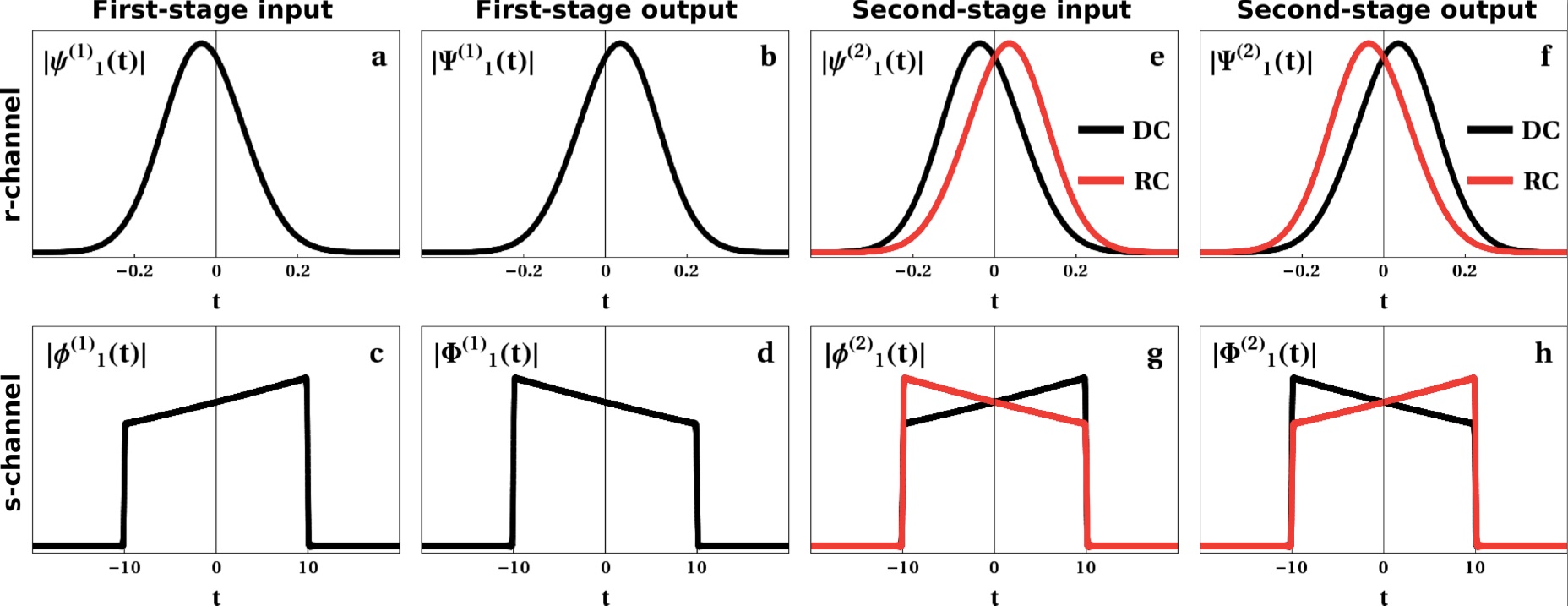}
\caption{ The dominant input (a, c, e, g) and output (b,
  d, f, h) Schmidt modes for the first-(a, b, c, d) and second-(e, f,
  g, h) stages for both {\it r} (a, b, e, f) and {\it s} (c, d, g, h)
  channels for both RC (red, lighter shade) and DC (black)
  configurations of TWM-TMI for a Gaussian-shaped pump, and
  $\zeta=|\beta'_r-\beta'_s|l/\tau_p = 200$. The values of $t$ are
  relative to a $|\beta'_r-\beta'_s|l$ of $20$. Due to the nature of
  temporal skewing, the inter-stage mode-matching between the
  first-stage output Schmidt modes (b, d) and the second-stage input
  Schmidt modes (e, g) is larger for the RC than the DC configuration,
  thus yielding better selectivity. The complete two-stage composite
  system Schmidt modes for TWM-TMI in the RC configuration are plotted
  in Fig. \ref{fig06}.}
\label{fig05}
\end{figure*}

\begin{figure*}[htb]
\centering
\includegraphics[width=\linewidth]{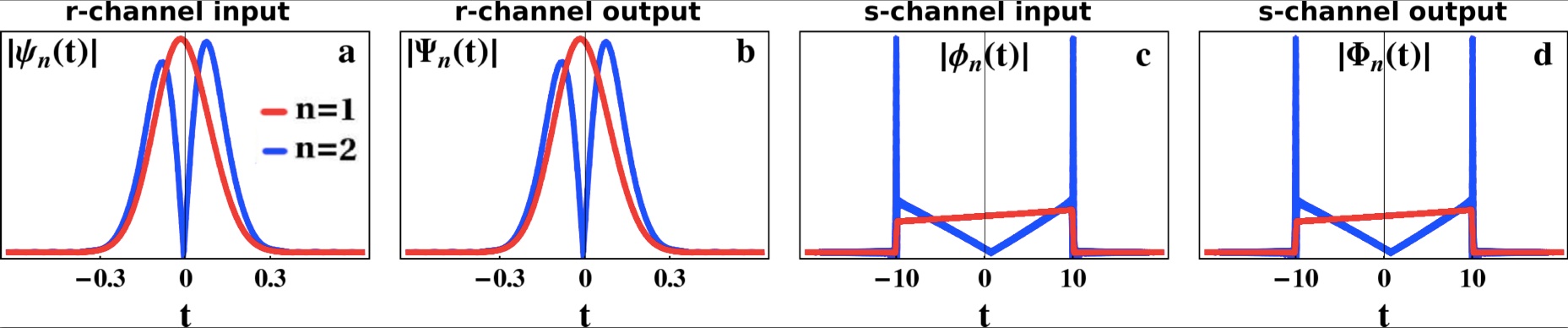}
\caption{ Schmidt modes for TWM-TMI in the RC
  configuration with Gaussian pump, and
  $\zeta=|\beta'_r-\beta'_s|l/\tau_p = 200$, yielding a selectivity of
  $0.9846$. The magnitudes of the first two Schmidt modes are shown
  for {\it r}-input (a), {\it r}-output (b), {\it s}-input (c), and
  {\it s}-output (d). Values of $t$ are relative to a
  $|\beta'_r-\beta'_s|l$ of $20$. Fig. \ref{fig07} shows the
  conversion efficiencies of the first four Schmidt modes for two
  different pump shapes.}
\label{fig06}
\end{figure*}

\begin{figure}[h!]
\centering
\includegraphics[width=\columnwidth]{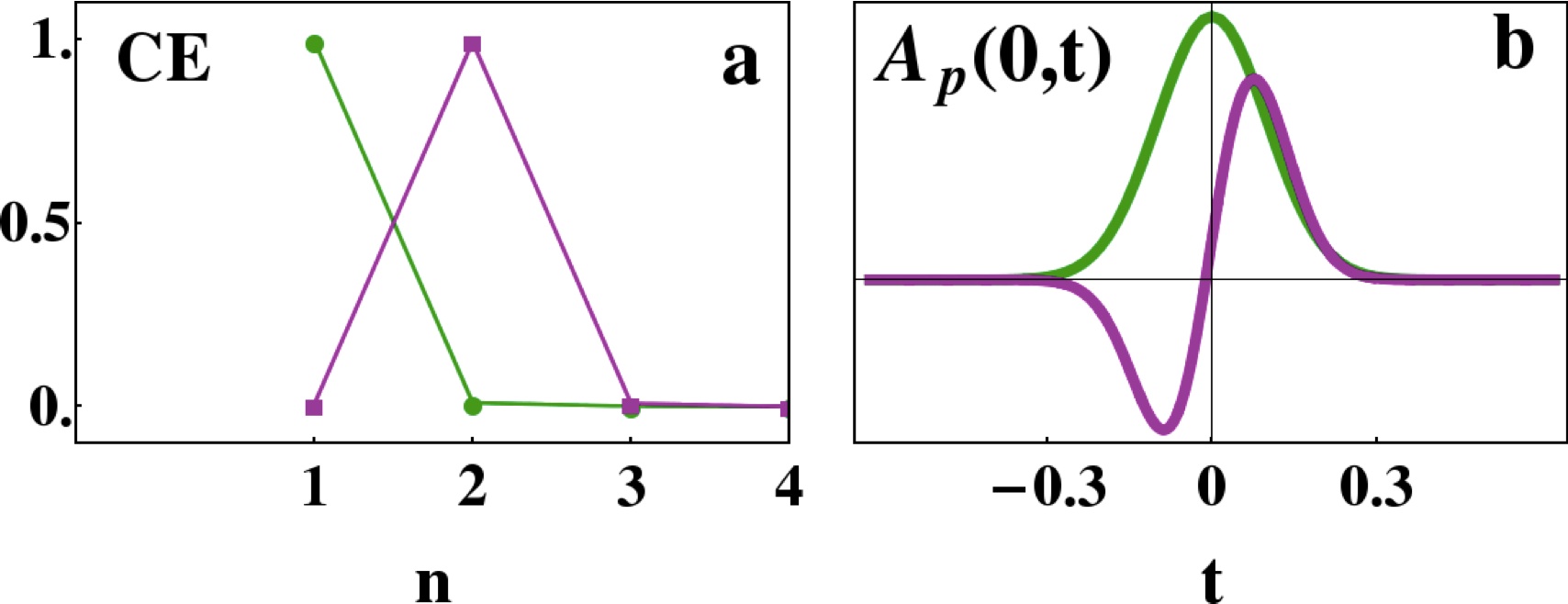}
\caption{ (a) Conversion efficiencies for
  Gaussian-pumped-TWM-TMI Schmidt modes in the RC configuration. The
  CE for a Gaussian pump are shown in green (darker shade), and those
  for a custom pump shape tailored to drop the second Schmidt mode are
  shown in magenta (lighter shade). (b) The corresponding pump shapes
  shown with matching colors. Values of $t$ are relative to a
  $|\beta'_r-\beta'_s|l$ of $20$.}
\label{fig07}
\end{figure}

Using ($A_r(0,t)=0$, $A_s(0,t')=\tilde{B}_{s,l}(t')$) as inputs for
the solver, and decomposing the resulting {\it r}-channel outputs
$A_r(l,t)$ in the $\{B_{r,k}\}$ basis, will yield the entire
$l$th column of the complex matrix $\overline{G}_{rs}$. Once this
matrix is determined, its Schmidt decomposition will directly reveal
$\{U_{jk}\}$, $\{V_{jl}\}$, and $\{\rho_j\}$, and through them, the
Schmidt modes. For all the results presented, we chose
Hermite-Gaussian functions for the spanning set of basis functions for
both input and output Schmidt modes, since the low-CE Schmidt modes
for Gaussian pump shapes are nearly Hermite-Gaussian
\cite{uren05,mej12}.

\subsubsection{Three-wave mixing}

Consider the case in which the target mode is $50$\% converted in each
stage individually while orthogonal modes remain almost completely
unconverted. Then, using Eqs. (\ref{svdeq}) in Eq. (\ref{eq15}) shows
the need for inter-stage temporal mode matching between the output
Schmidt modes of the first stage and the input Schmidt modes of the
second stage for the scheme to work. Figure \ref{fig05} shows the
plots for the input and output Schmidt modes of both stages for both
RC and DC configuration TMI, with a Gaussian pump, and the parameter
$\zeta=|\beta'_r-\beta'_s|l/\tau_p = 200$, where $l$ is the per stage
medium length and $\tau_p$ is the Gaussian pump width. The Schmidt
modes of a single channel for a given stage are found to be temporally
symmetrically skewed in opposite directions between the input and the
output, the direction being dependent on the sign of
$(\beta'_r-\beta'_s)$. Consequently, the inter-stage mode matching
will be exact for the RC configuration, and inferior for the DC
configuration. For $\zeta = 200$, the RC
$\mu^{TWM}_{1,1}=\eta^{TWM}_{1,1}=1$. But for DC,
$\mu^{TWM}_{1,1}=0.983$ and $\eta^{TWM}_{1,1}=0.901$.

Figures \ref{fig06}(a-d) Schmidt modes for the RC configuration
TWM-TMI for $\zeta= 200$ and a Gaussian pump. The selectivity {\it S}
was computed to be $0.9846$ ($|\rho_1|^2=0.9975$,
$|\rho_2|^2=0.0110$). $\zeta$ is also the ratio of time-widths of the
{\it r}- and {\it s}-Schmidt modes. The CE of the first four Schmidt
modes ({\it i.e.} $|\rho_n|^2$) are plotted in green in
Fig. \ref{fig07}(a), along with the CE of the same Schmidt modes when
using a custom pump shape (magenta) tailored to `drop' the second {\it
  r}-input Schmidt mode. Since the TWM pump can influence only the
shape of the {\it r}-channel Schmidt modes, the tailored pump in
Fig. \ref{fig07}(b) will frequency convert the second {\it r}-input
Schmidt mode from Fig. \ref{fig06}(a) into an {\it s}-channel TM
identical to the first {\it s}-output Schmidt mode from
Fig. \ref{fig06}(d).

As an example, with a Gaussian pump, a signal photon in the {\it
  r} channel with shape $\psi_1(t)$ (Fig. \ref{fig06}(a)) will be
frequency converted into an {\it s}-channel photon with shape
$\Phi_1(t)$ (Fig. \ref{fig06}(d)), with an efficiency of
$|\rho_1|^2=0.9975$. In contrast, a photon in the {\it r} channel with
shape $\psi_2(t)$ will be frequency converted into an {\it s}-channel
photon with shape $\Phi_2(t)$, with a very small efficiency of
$|\rho_2|^2=0.0110$. In other words, an {\it r}-channel photon stays
in the {\it r} channel with probability $0.9890$, and will exit the
device with shape $\Psi_2(t)$ (Fig. \ref{fig06}(b)). Figure
\ref{fig06}(c) comes into play only if the input photon is in the {\it
  s} channel.

For the DC configuration with $\zeta=200$, we computed a selectivity
of $0.9805$ (with $|\rho_1|^2=0.9957$, $|\rho_2|^2=0.0134$). This is
slightly lower than the RC configuration due to relatively inferior
inter-stage mode-matching. 

\begin{figure}[htb]
\centering
\includegraphics[width=\columnwidth]{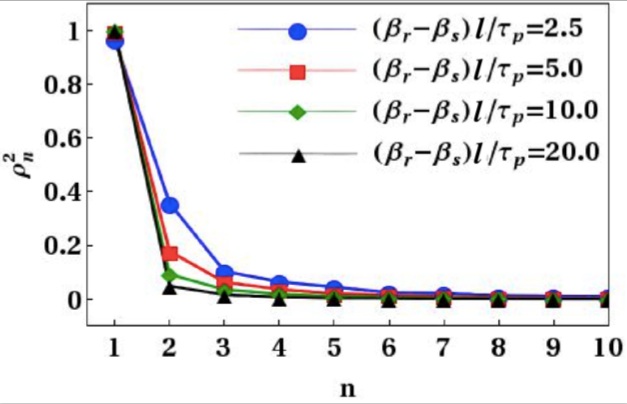}
\caption{ Dominant Schmidt mode conversion efficiencies
  for TWM-TMI in the RC configuration, illustrating asymptotic
  improvement in selectivity with decreasing pump-pulse width $\tau_p$
  relative to interaction time $(\beta'_r-\beta'_s)l$.}
\label{fig08}
\end{figure}

\begin{figure}[htb]
\centering
\includegraphics[width=\columnwidth]{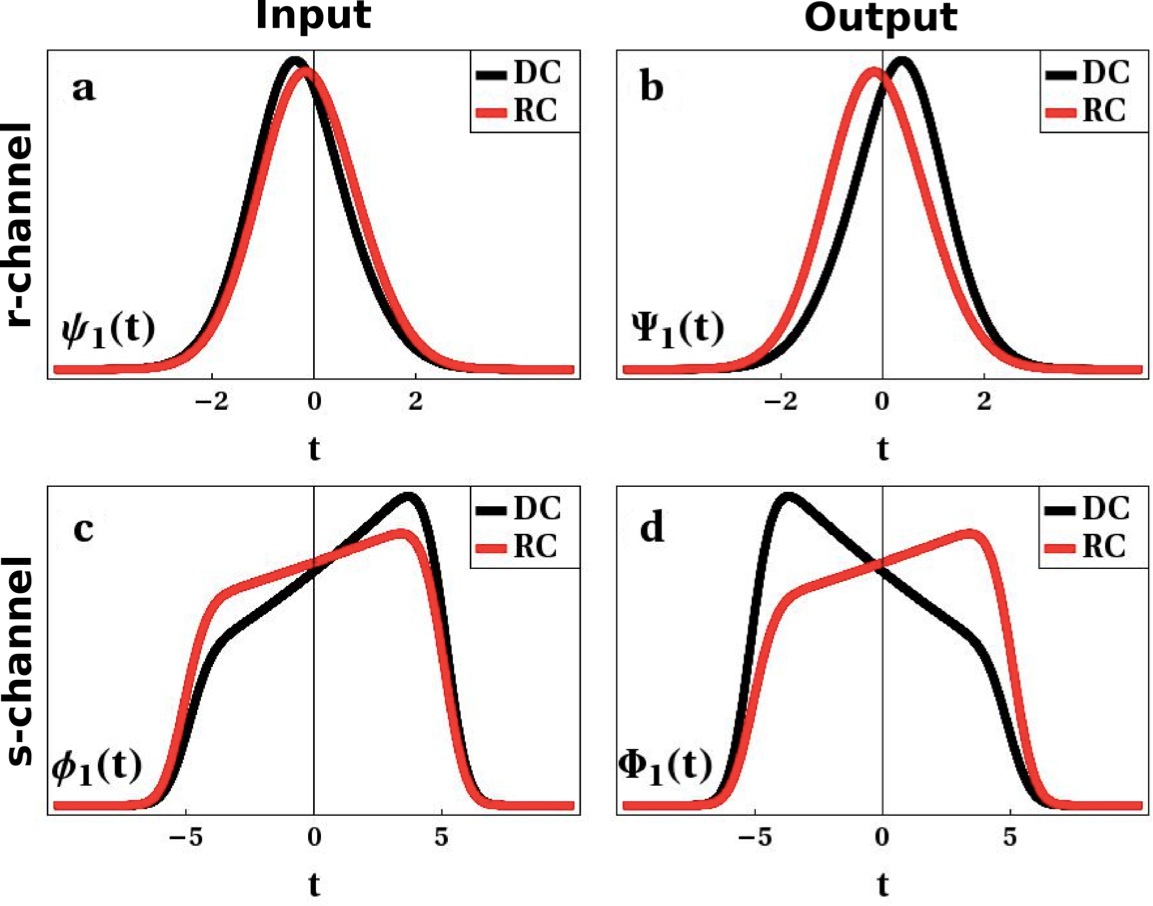}
\caption{ The first Schmidt modes of (a) {\it r}-input,
  (b) {\it r}-output, (c) {\it s}-input, and (d) {\it s}-output
  channels for both DC (black) and RC (red, lighter shade)
  configurations of a two-stage TMI implemented using TWM with a
  single Gaussian pump, and $|\beta'_r-\beta'_s|l/\tau_p=10$. Values
  of $t$ are relative to a $|\beta'_r-\beta'_s|l$ of $10$.}
\label{fig09}
\end{figure}

Both single-stage Green function separability, and inter-stage mode
matching improve asymptotically with increasing $\zeta$, providing for
corresponding gains in selectivity (Fig. \ref{fig08}). Larger $\zeta$
also implies a decrease in temporal skewness of the Schmidt modes
relative to the pump shapes. The exact mode matching in RC
configuration results in matching skewness directions for input and
output Schmidt modes of a single channel, a feature that is not
present for DC configuration (Fig. \ref{fig09}). Therefore,
unconverted higher-order Schmidt modes of the RC process undergo no
temporal distortion upon passing through the TMI transformation. This
allows TWM-TMI devices operating in the RC configuration to be used in
a chained sequence to implement multiple operations on the
temporal-mode basis.

TMI is physically easier to implement in the DC configuration. A
$\zeta$ of $200$ can be realized, for example, in a typical $5$-cm
long periodically poled lithium niobate waveguide with a $70$-fs
pump-pulse and signal wavelengths at $795$ nm and $1324$ nm. RC
configuration implementations rely on media with negative dispersion,
which exist for $\chi^{(3)}$-media in the form of photonic-crystal
fibers, and are also possible to engineer in $\chi^{(2)}$-media using
photonic crystal waveguides \cite{drid00}.

\begin{figure}[htb]
\centering
\includegraphics[width=0.95\columnwidth]{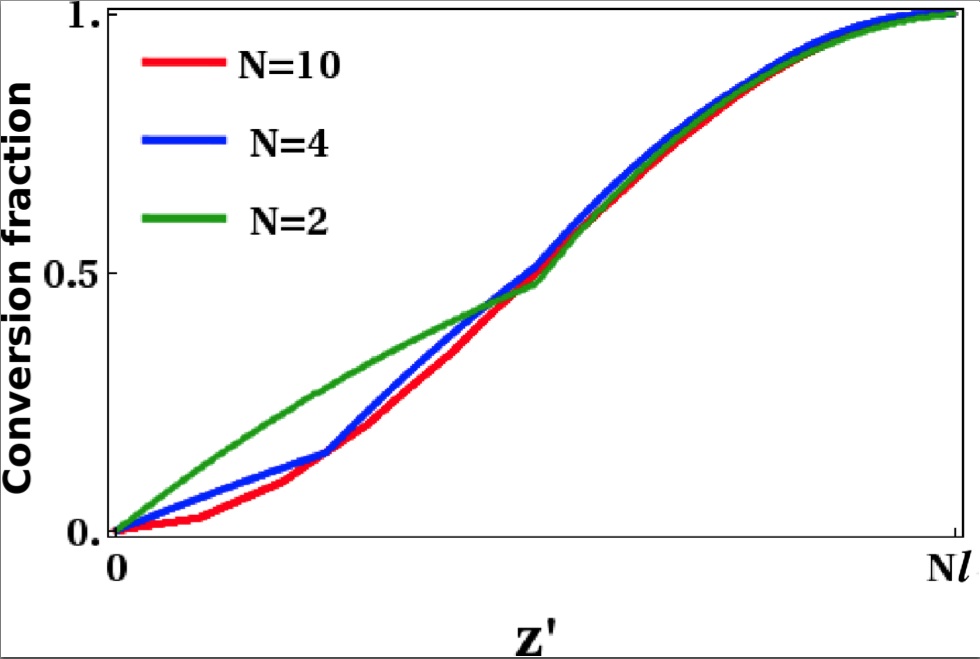}
\caption{ Multistage TWM implementation of TMI, showing
  the ratio of converted to total energy of the first Schmidt mode
  versus $z'$ (propagation distance). $N$ is the number of stages and
  $l$ is the medium length for each stage.}
\label{fig10}
\end{figure}
\begin{figure}[htb]
\centering
\includegraphics[width=\columnwidth]{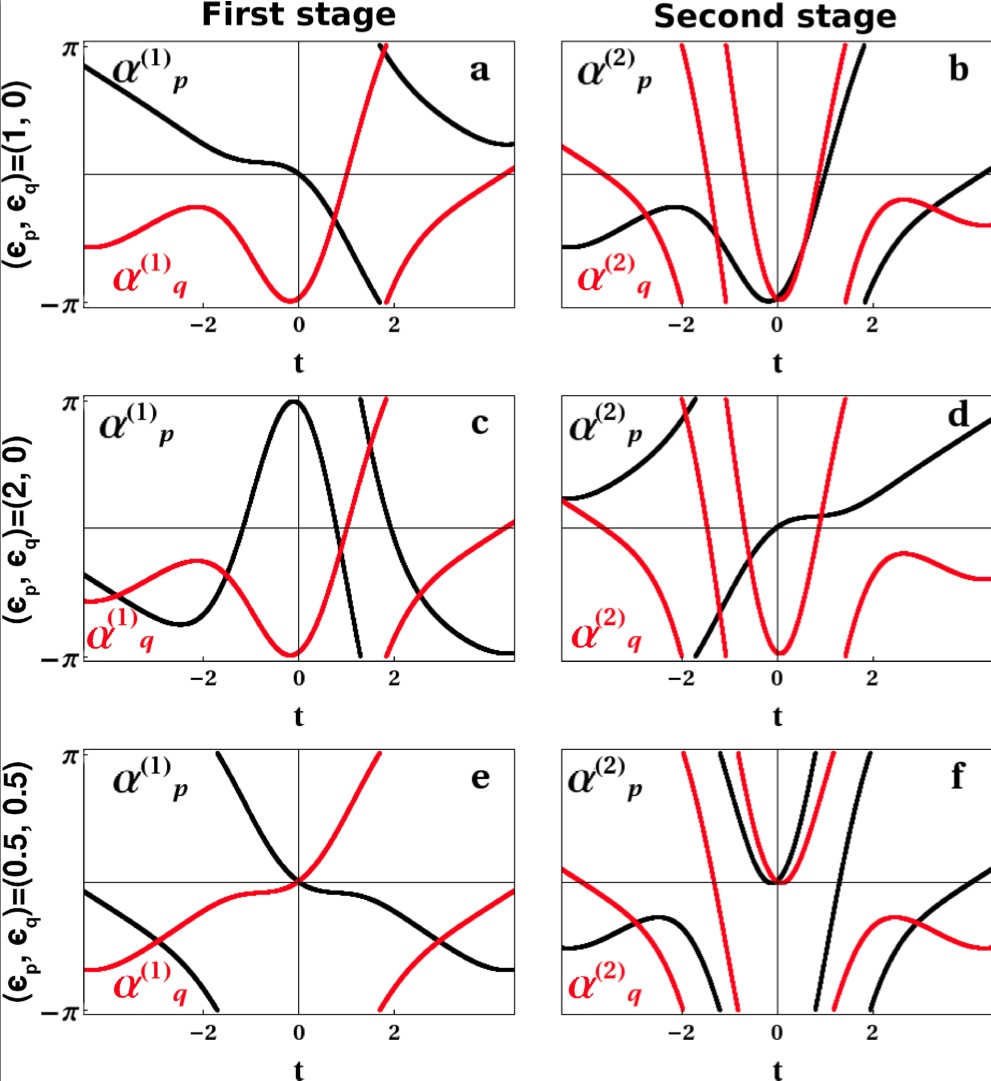}
\caption{ The first-stage (a, c, e) and second-stage (b,
  d, f) pump chirps used to derive the three inter-stage mode-matched
  Schmid-mode phase profiles shown in Fig. \ref{fig12}. The
  chirp-parameters \{$\epsilon_p$, $\epsilon_q$\} are \{1, 0\} in (a,
  b) , \{2, 0\} in (c, d) , and \{0.5, 0.5\} in (e, f). Values of $t$
  are relative to a $|\beta'_r-\beta'_s|l$ of $10$.}
\label{fig11}
\end{figure}

TMI can also be extended to arbitrarily large number of stages in both
RC and DC configurations. Since the single-stage Green functions are
more separable at lower conversion efficiencies, increasing the number
of stages can increase selectivity. Multistage implementations will
use lower pump powers due to the increase in the number of inter-pulse
interactions. For optimal selectivity when using an {\it N}-stage
process, the conversion efficiency for the first stage should be
approximately $0.5[1-\cos(\pi/N)]$. The expression is the exact
splitting ratio necessary for a sequence of $N$ nonpolarizing beam
splitters to interferometrically change the propagation direction of
an incident beam of coherent light, and becomes more accurate for TMI
for larger $N$.  Figure \ref{fig10} shows the ratio of
converted-to-total energy of the first Schmidt mode versus propagation
distance for various number of stages. The plots tend to
asymptotically converge to a sinusoidal curve. A four-stage scheme
yielded DC and RC selectivities of $0.9977$ and $0.9978$
respectively. The corresponding ten-stage selectivities are $0.99996$
and $0.99997$. Increasing the number of stages also decreases the
temporal skewness of the first Schmidt mode relative to the
corresponding pump-shape. The directionality of skewness for $N$-stage
TWM-TMI in the DC configuration is independent of $N$. However, for
$N$-stage TWM-TMI in the RC configuration, one can choose between
distortionless FC and distortionless unconverted transmission of
pulses by choosing odd or even $N$ respectively.

\begin{figure*}[htb]
\centering
\includegraphics[width=\linewidth]{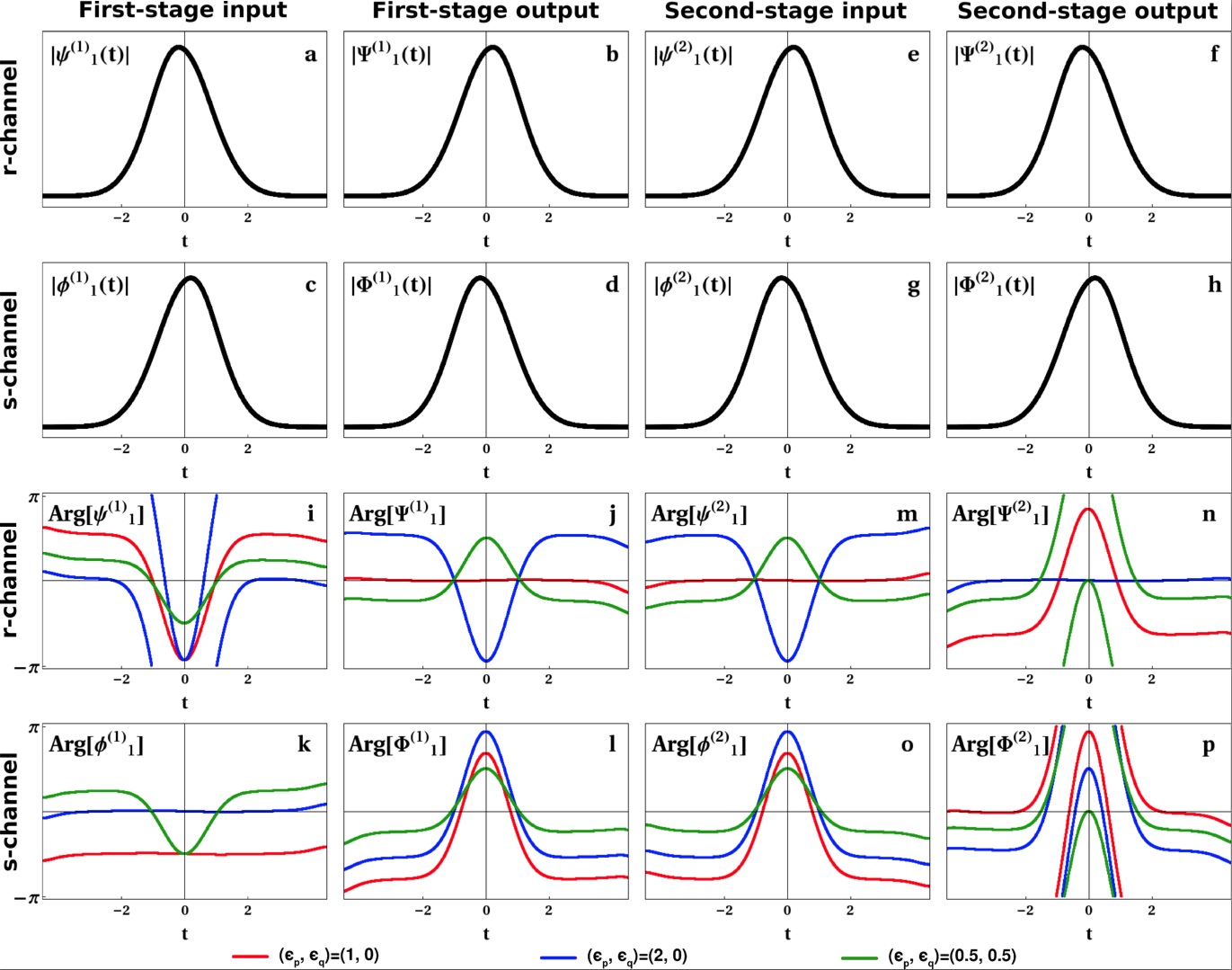}
\caption{ The dominant input (a, c, e, g) and output
  (b, d, f, h) Schmidt modes for the first-(a, b, c, d) and second-(e,
  f, g, h) stages for both {\it r} (a, b, e, f) and {\it s} (c, d, g,
  h) channels for RC configuration of FWM-TMI for a Gaussian-shaped
  pumps, and complete pump collisions
  ($|\beta'_r-\beta'_s|l/(\tau_p+\tau_q)=5$). Also shown are the
  corresponding Schmidt mode phase profiles (i-p) for three seperate
  choices of pump chirps for the two stages. Due to pump-induced self-
  and cross-phase modulation, the second-stage pump chirps have to be
  related to the first-stage pump chirps in the manner specified in
  Eq. \ref{eq13} for good inter-stage mode-matching between (b, d, j,
  l) and (e, g, m, o) respectively. The complete two-stage composite
  system Schmidt modes for FWM-TMI in the RC configuration are plotted
  in Fig. \ref{fig13}. Values of $t$ are relative to a
  $|\beta'_r-\beta'_s|l$ of $10$.}
\label{fig12}
\end{figure*}

\subsubsection{Four-wave mixing}

In FWM, there are two strong pulsed pumps, each of which needs to be
group-velocity-matched to a different signal channel to ensure good
single-stage Green function separability. In addition, the
$\chi^{(3)}$ medium being used in each stage must be long enough for
complete inter-pump-pulse collision to occur. This is straightforward
to satisfy with the use of highly-nonlinear optical fibers. For
Gaussian pumps of temporal widths of the order of $10$ ps, and
wavelengths at $800$ nm and $850$ nm, a typical highly-nonlinear
photonic crystal fiber will need to be about $20$ m long for complete
collision ($|\beta'_r-\beta'_s|l/(\tau_p+\tau_q)=5$).

The presence of two pumps enables independent shaping of the Schmidt
modes of both channels in FWM-TMI. However, nonlinear phase modulation
(Eq. \ref{eomfwm}) severely affects the Schmidt-mode phase-profiles,
and restricts FWM-TMI to the RC configuration. This is a result of the
well-known cross- and self-phase modulation present in any
$\chi^{(3)}$ medium. To overcome this impairment, the pumps must be
pre-chirped with specific phase-profiles \cite{mej12b} for each stage
to enhance inter-stage mode-matching. The starting phase profiles of
the pump {\it p} and {\it q} in stage-$\xi$ ( $\alpha^{(\xi)}_p(t)$
and $\alpha^{(\xi)}_q(t)$ respectively) need to be

\begin{widetext}
\begin{align}
&\alpha^{(\xi)}_p(t)=-2\overline\gamma^{(\xi)}\int\limits^t_{-\infty}\left[|\overline{A_q}(s)|^2-|\overline{A_p}(s)|^2\right]ds-\frac{3}{2}\overline\gamma^{(\xi)}|\overline{A_p}(t)|^2t+\frac{3}{2}\overline\gamma^{(\xi)}|\overline{A_q}(0)|^2t+(\epsilon_p-\delta_{\xi,2})\gamma l |\overline{A_p}(t)|^2,\notag\\ 
&\alpha^{(\xi)}_q(t)=-2\overline\gamma^{(\xi)}\int\limits^t_{-\infty}\left[|\overline{A_q}(s)|^2-|\overline{A_p}(s)|^2\right]ds+\frac{3}{2}\overline\gamma^{(\xi)}|\overline{A_q}(t)|^2t-\frac{3}{2}\overline\gamma^{(\xi)}|\overline{A_p}(0)|^2t+(\epsilon_q-\delta_{\xi,2})\gamma l |\overline{A_q}(t)|^2,\label{eq13}
\end{align}
\end{widetext}

\noindent where $\overline\gamma^{(\xi)} =
\gamma/(\beta'^{(\xi)}_r-\beta'^{(\xi)}_s)$, and
$\overline{A_j}(t)\equiv A_j(0,t)$. The limits of the integration
terms are valid only for complete pump collisions. It must be noted
that the pump phase profiles contain a term each that is fully linear
in time, which are equivalent to frequency shifts. These are an
approximate compensation for a non-separable part of the phase profile
of the GF kernels \cite{mej12b} which only attains significant
magnitude away from the centroids of the GF-kernel amplitude
functions. The slope of this term for each pump is proportional to the
magnitude-square of the amplitude of the other pump's envelope. The
slope is also proportional to $\overline\gamma^{(\xi)}$, which changes
sign between the two stages. For a given set of chirp parameters, a
pump will require frequency shifts of opposite signs in the two
stages. Figure \ref{fig11} shows the plots of the pump phase profiles
from Eq. \ref{eq13} for three sets of chirp parameter values. Figure
\ref{fig12} shows the amplitudes and phase profiles of the first input
and first output Schmidt modes for both stages of FWM-TMI in the RC
configuration for the same three sets of chirp parameter values.

\begin{figure*}[htb]
\centering
\includegraphics[width=\linewidth]{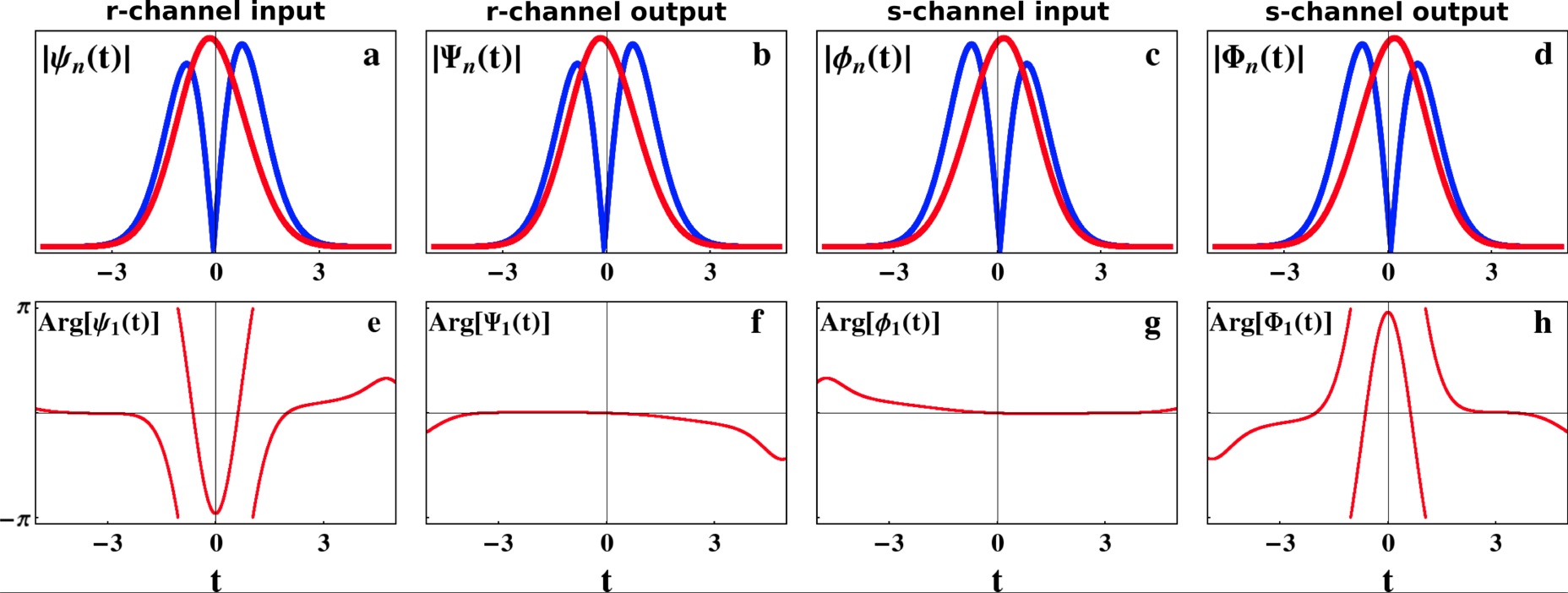}
\caption{ The first two Schmidt modes (a, b, c, d), and
  first Schmidt-mode phase profiles (e, f, g, h) for FWM-TMI in the RC
  configuration with Gaussian pumps. The first two Schmidt modes are
  shown for {\it r}-input (a), {\it r}-output (b), {\it s}-input (c),
  and {\it s}-output (d). Pump pre-chirp parameters used were
  $\epsilon_p = 2$, $\epsilon_q = 0$. The selectivity was
  $0.9873$. Values of $t$ are relative to a $|\beta'_r-\beta'_s|l$ of
  $10$.}
\label{fig13}
\end{figure*}

The chirp parameters $\epsilon_p$ and $\epsilon_q$ can be any real
values. However, for FWM-TMI in the RC configuration to be useful, one
will have to pick specific values to make the desired channel
input/output Schmidt mode have a flat phase profile. For example,
Figs. \ref{fig13}(a)-\ref{fig13}(d) show the Schmidt-mode amplitudes,
and phases corresponding to FWM-TMI with selectivity $0.9873$
($|\rho_1|^2=0.9973$, $|\rho_2|^2=0.0082$). The pump pre-chirps
(Fig. \ref{fig11}(c, d)) were specifically chosen to yield flat phase
profiles for the {\it r}-output (Fig. \ref{fig13}(f)) and {\it
  s}-input (Fig. \ref{fig13}(g)) Schmidt modes. Namely, $\epsilon_p=2$
and $\epsilon_q=0$. The results were computed for
$|\beta'_r-\beta'_s|l/(\tau_p+\tau_q)=5$.

 FWM-TMI can in principle be extended to multiple stages, provided
 that every stage-interface is in the RC configuration. The pump
 frequency shift will alternate in sign for every stage in
 sequence. The pump chirp profile functions in Eq. \ref{eq13} will
 likewise need to be generalized for arbitrary number of stages. The
 Kronecker-$\delta_{\xi,2}$ term will pick up a factor inversely
 proportional to the number of stages.

The need for pump-phase chirps in FWM-TMI can be circumvented if one
resorts to asymmetrically-pumped Bragg scattering \cite{mej14}, in
which one of the pumps is made very long or continuous wave (CW), and
the shorter pump power is relatively weak. Then, all self- and
cross-phase modulation effects occur only due to the long CW pump,
which is equivalent to pump-power-dependent frequency and wavenumber
shifts of the signals. This method can simulate TWM-like dynamics in a
$\chi^{(3)}$ medium, and by extension, can be used to implement
TWM-TMI using FWM. Pump-signal group-velocity matching is easier to
achieve in $\chi^{(3)}$ media with waves whose interaction is phase
matched, and the choice of which pump is to be made CW does not affect
the selectivity, making asymmetrically pumped Bragg scattering a
valuable option warranting further study.

\section{V. Concluding remarks}

TMI can nearly perfectly separate field-orthogonal temporal modes into
separate beams corresponding to different carrier frequencies, as well
as combine such TMs into a single beam. It also enables real-time
routing capabilities in all-optical networks while transcending the
limitations of narrow-band filtering and projective measurement
schemes \cite{hayat}. It offers true field-orthogonal realizations of
both classical \cite{herit} and quantum \cite{hayat} optical
code-division multiple-access (OCDMA) networks with high spectral
efficiency for both channel routing and detection, unhindered by
multiuser interference.

Breaking the $80$\% single-stage selectivity limit provides a means of
physically realising high-fidelity quantum-unitary devices such as the
proposed `quantum pulse gate' (orthogonal-field TM sorter) and
`quantum pulse shaper' \cite{brecht11}, which operate explicitly on
the TM Hilbert-space. TMI can aid in long-distance quantum
communication and quantum key distribution using novel
continuous-variable encodings of qubits or qudits in the
time-frequency basis \cite{nunn13}. The tunable interferometric
phase-angle allows TMI to simulate linear optical elements analogous
to the action of polarizing and non-polarizing beam splitters,
half-wave plates and quarter-wave plates on the polarization
basis. TMI devices can therefore perform arbitrary unitary
transformations on the TM space, and in conjunction with multimode
quantum memories (such as atomic Raman memories
\cite{polzik11,humph14}), open a new avenue for linear optical quantum
computing in a single spatial mode \cite{humph13}. All of these
applications will be ellaborated upon in future publications.

In conclusion, TMI can selectively manipulate photonic temporal mode
components of any shape accessible by our ability to reshape strong
pump pulses \cite{cundiff10,weiner11}. Modular TMI devices also
possess a phase control ($\theta$) that can suppress QFC completely
and on demand. TMI may be implemented using either three-wave mixing,
or four-wave mixing enabled frequency conversion, and may be
generalized to multiple stages for gains in selectivity asymptotically
approaching unity. The addition of temporal-mode analysis to
polarization and orbital angular momentum analysis completes the
toolkit for encoding information on quantum and classical states of
light.

We thank L. Mejling, K. Rottwitt and J. Nunn for helpful
discussions. This work was supported by the National Science
Foundation through EPMD and GOALI, grant ECCS-1101811.




\begin{thebibliography}{50}%
\makeatletter
\providecommand \@ifxundefined [1]{%
 \@ifx{#1\undefined}
}%
\providecommand \@ifnum [1]{%
 \ifnum #1\expandafter \@firstoftwo
 \else \expandafter \@secondoftwo
 \fi
}%
\providecommand \@ifx [1]{%
 \ifx #1\expandafter \@firstoftwo
 \else \expandafter \@secondoftwo
 \fi
}%
\providecommand \natexlab [1]{#1}%
\providecommand \enquote  [1]{``#1''}%
\providecommand \bibnamefont  [1]{#1}%
\providecommand \bibfnamefont [1]{#1}%
\providecommand \citenamefont [1]{#1}%
\providecommand \href@noop [0]{\@secondoftwo}%
\providecommand \href [0]{\begingroup \@sanitize@url \@href}%
\providecommand \@href[1]{\@@startlink{#1}\@@href}%
\providecommand \@@href[1]{\endgroup#1\@@endlink}%
\providecommand \@sanitize@url [0]{\catcode `\\12\catcode `\$12\catcode
  `\&12\catcode `\#12\catcode `\^12\catcode `\_12\catcode `\%12\relax}%
\providecommand \@@startlink[1]{}%
\providecommand \@@endlink[0]{}%
\providecommand \url  [0]{\begingroup\@sanitize@url \@url }%
\providecommand \@url [1]{\endgroup\@href {#1}{\urlprefix }}%
\providecommand \urlprefix  [0]{URL }%
\providecommand \Eprint [0]{\href }%
\providecommand \doibase [0]{http://dx.doi.org/}%
\providecommand \selectlanguage [0]{\@gobble}%
\providecommand \bibinfo  [0]{\@secondoftwo}%
\providecommand \bibfield  [0]{\@secondoftwo}%
\providecommand \translation [1]{[#1]}%
\providecommand \BibitemOpen [0]{}%
\providecommand \bibitemStop [0]{}%
\providecommand \bibitemNoStop [0]{.\EOS\space}%
\providecommand \EOS [0]{\spacefactor3000\relax}%
\providecommand \BibitemShut  [1]{\csname bibitem#1\endcsname}%
\let\auto@bib@innerbib\@empty
\bibitem [{\citenamefont {Smith}\ and\ \citenamefont {Raymer}(2007)}]{smi07}%
  \BibitemOpen
  \bibfield  {author} {\bibinfo {author} {\bibfnamefont {B.~J.}\ \bibnamefont
  {Smith}}\ and\ \bibinfo {author} {\bibfnamefont {M.~G.}\ \bibnamefont
  {Raymer}},\ }\href {\doibase 10.1088/1367-2630/9/11/414} {\bibfield
  {journal} {\bibinfo  {journal} {New J. Phys.}\ }\textbf {\bibinfo {volume}
  {9}},\ \bibinfo {pages} {414} (\bibinfo {year} {2007})}\BibitemShut {NoStop}%
\bibitem [{\citenamefont {Sasada}\ and\ \citenamefont {Okamoto}(2003)}]{sas03}%
  \BibitemOpen
  \bibfield  {author} {\bibinfo {author} {\bibfnamefont {H.}~\bibnamefont
  {Sasada}}\ and\ \bibinfo {author} {\bibfnamefont {M.}~\bibnamefont
  {Okamoto}},\ }\href {\doibase 10.1103/PhysRevA.68.012323} {\bibfield
  {journal} {\bibinfo  {journal} {Phys. Rev. A}\ }\textbf {\bibinfo {volume}
  {68}},\ \bibinfo {pages} {012323} (\bibinfo {year} {2003})}\BibitemShut
  {NoStop}%
\bibitem [{\citenamefont {Yarnall}\ \emph {et~al.}(2007)\citenamefont
  {Yarnall}, \citenamefont {Abouraddy}, \citenamefont {Saleh},\ and\
  \citenamefont {Teich}}]{abou07}%
  \BibitemOpen
  \bibfield  {author} {\bibinfo {author} {\bibfnamefont {T.}~\bibnamefont
  {Yarnall}}, \bibinfo {author} {\bibfnamefont {A.~F.}\ \bibnamefont
  {Abouraddy}}, \bibinfo {author} {\bibfnamefont {B.~E.~A.}\ \bibnamefont
  {Saleh}}, \ and\ \bibinfo {author} {\bibfnamefont {M.~C.}\ \bibnamefont
  {Teich}},\ }\href {\doibase 10.1103/PhysRevLett.99.250502} {\bibfield
  {journal} {\bibinfo  {journal} {Phys. Rev. Lett.}\ }\textbf {\bibinfo
  {volume} {99}},\ \bibinfo {pages} {250502} (\bibinfo {year}
  {2007})}\BibitemShut {NoStop}%
\bibitem [{\citenamefont {Winzer}(2009)}]{winzer09}%
  \BibitemOpen
  \bibfield  {author} {\bibinfo {author} {\bibfnamefont {P.~J.}\ \bibnamefont
  {Winzer}},\ }\href {http://photonicssociety.org/newsletters/feb09/} {\enquote
  {\bibinfo {title} {Modulation and multiplexing in optical communication
  systems},}\ }\bibinfo {howpublished} {{IEEE LEOS Newsletter, {\bf 23} (1), 4
  }} (\bibinfo {year} {2009})\BibitemShut {NoStop}%
\bibitem [{\citenamefont {Barreiro}\ \emph {et~al.}(2008)\citenamefont
  {Barreiro}, \citenamefont {Wei},\ and\ \citenamefont {Kwiat}}]{bar10}%
  \BibitemOpen
  \bibfield  {author} {\bibinfo {author} {\bibfnamefont {J.~T.}\ \bibnamefont
  {Barreiro}}, \bibinfo {author} {\bibfnamefont {T.~C.}\ \bibnamefont {Wei}}, \
  and\ \bibinfo {author} {\bibfnamefont {P.~G.}\ \bibnamefont {Kwiat}},\
  }\href@noop {} {\bibfield  {journal} {\bibinfo  {journal} {Nature Phys.}\
  }\textbf {\bibinfo {volume} {4}},\ \bibinfo {pages} {282} (\bibinfo {year}
  {2008})}\BibitemShut {NoStop}%
\bibitem [{\citenamefont {Berkhout}\ \emph {et~al.}(2010)\citenamefont
  {Berkhout}, \citenamefont {Lavery}, \citenamefont {Courtial}, \citenamefont
  {Beijersbergen},\ and\ \citenamefont {Padgett}}]{pad10}%
  \BibitemOpen
  \bibfield  {author} {\bibinfo {author} {\bibfnamefont {G.~C.~G.}\
  \bibnamefont {Berkhout}}, \bibinfo {author} {\bibfnamefont {M.~P.~J.}\
  \bibnamefont {Lavery}}, \bibinfo {author} {\bibfnamefont {J.}~\bibnamefont
  {Courtial}}, \bibinfo {author} {\bibfnamefont {M.~W.}\ \bibnamefont
  {Beijersbergen}}, \ and\ \bibinfo {author} {\bibfnamefont {M.~J.}\
  \bibnamefont {Padgett}},\ }\href {\doibase 10.1103/PhysRevLett.105.153601}
  {\bibfield  {journal} {\bibinfo  {journal} {Phys. Rev. Lett.}\ }\textbf
  {\bibinfo {volume} {105}},\ \bibinfo {pages} {153601} (\bibinfo {year}
  {2010})}\BibitemShut {NoStop}%
\bibitem [{\citenamefont {Yao}\ and\ \citenamefont {Padgett}(2011)}]{yao11}%
  \BibitemOpen
  \bibfield  {author} {\bibinfo {author} {\bibfnamefont {A.~M.}\ \bibnamefont
  {Yao}}\ and\ \bibinfo {author} {\bibfnamefont {M.~J.}\ \bibnamefont
  {Padgett}},\ }\href {\doibase 10.1364/AOP.3.000161} {\bibfield  {journal}
  {\bibinfo  {journal} {Adv. Opt. Photon.}\ }\textbf {\bibinfo {volume} {3}},\
  \bibinfo {pages} {161} (\bibinfo {year} {2011})}\BibitemShut {NoStop}%
\bibitem [{\citenamefont {Wang}\ \emph {et~al.}(2012)\citenamefont {Wang},
  \citenamefont {Yang}, \citenamefont {Fazal}, \citenamefont {Ahmed},
  \citenamefont {Yan}, \citenamefont {Huang}, \citenamefont {Ren},
  \citenamefont {Yue}, \citenamefont {Dolinar}, \citenamefont {Tur},\ and\
  \citenamefont {Willner}}]{wang12}%
  \BibitemOpen
  \bibfield  {author} {\bibinfo {author} {\bibfnamefont {J.}~\bibnamefont
  {Wang}}, \bibinfo {author} {\bibfnamefont {J.-Y.}\ \bibnamefont {Yang}},
  \bibinfo {author} {\bibfnamefont {I.~M.}\ \bibnamefont {Fazal}}, \bibinfo
  {author} {\bibfnamefont {N.}~\bibnamefont {Ahmed}}, \bibinfo {author}
  {\bibfnamefont {Y.}~\bibnamefont {Yan}}, \bibinfo {author} {\bibfnamefont
  {H.}~\bibnamefont {Huang}}, \bibinfo {author} {\bibfnamefont
  {Y.}~\bibnamefont {Ren}}, \bibinfo {author} {\bibfnamefont {Y.}~\bibnamefont
  {Yue}}, \bibinfo {author} {\bibfnamefont {S.}~\bibnamefont {Dolinar}},
  \bibinfo {author} {\bibfnamefont {M.}~\bibnamefont {Tur}}, \ and\ \bibinfo
  {author} {\bibfnamefont {A.~E.}\ \bibnamefont {Willner}},\ }\href {\doibase
  nphoton.2012.138} {\bibfield  {journal} {\bibinfo  {journal} {Nature
  Photon.}\ }\textbf {\bibinfo {volume} {6}},\ \bibinfo {pages} {488}
  (\bibinfo {year} {2012})}\BibitemShut {NoStop}%
\bibitem [{\citenamefont {Bozinovic}\ \emph {et~al.}(2013)\citenamefont
  {Bozinovic}, \citenamefont {Yue}, \citenamefont {Ren}, \citenamefont {Tur},
  \citenamefont {Kristensen}, \citenamefont {Huang}, \citenamefont {Willner},\
  and\ \citenamefont {Ramachandran}}]{wil13}%
  \BibitemOpen
  \bibfield  {author} {\bibinfo {author} {\bibfnamefont {N.}~\bibnamefont
  {Bozinovic}}, \bibinfo {author} {\bibfnamefont {Y.}~\bibnamefont {Yue}},
  \bibinfo {author} {\bibfnamefont {Y.}~\bibnamefont {Ren}}, \bibinfo {author}
  {\bibfnamefont {M.}~\bibnamefont {Tur}}, \bibinfo {author} {\bibfnamefont
  {P.}~\bibnamefont {Kristensen}}, \bibinfo {author} {\bibfnamefont
  {H.}~\bibnamefont {Huang}}, \bibinfo {author} {\bibfnamefont {A.~E.}\
  \bibnamefont {Willner}}, \ and\ \bibinfo {author} {\bibfnamefont
  {S.}~\bibnamefont {Ramachandran}},\ }\href {\doibase 10.1126/science.1237861}
  {\bibfield  {journal} {\bibinfo  {journal} {Science}\ }\textbf {\bibinfo
  {volume} {340}},\ \bibinfo {pages} {1545} (\bibinfo {year}
  {2013})}\BibitemShut {NoStop}%
\bibitem [{\citenamefont {Yang}\ \emph {et~al.}(2011)\citenamefont {Yang},
  \citenamefont {Amin},\ and\ \citenamefont {Shieh}}]{yang11}%
  \BibitemOpen
  \bibfield  {author} {\bibinfo {author} {\bibfnamefont {Q.}~\bibnamefont
  {Yang}}, \bibinfo {author} {\bibfnamefont {A.~A.}\ \bibnamefont {Amin}}, \
  and\ \bibinfo {author} {\bibfnamefont {W.}~\bibnamefont {Shieh}},\ }in\
  \href@noop {} {\emph {\bibinfo {booktitle} {Impact of nonlinearities on
  Fiber Optic Communications}}},\ \bibinfo {editor} {edited by\ \bibinfo
  {editor} {\bibfnamefont {S.}~\bibnamefont {Kumar}}}\ (\bibinfo  {publisher}
  {Springer},\ \bibinfo {year} {2011}), ch. 2\BibitemShut {NoStop}%
\bibitem [{\citenamefont {Nakazawa}\ \emph {et~al.}(2014)\citenamefont
  {Nakazawa}, \citenamefont {Yoshida},\ and\ \citenamefont
  {Hirooka}}]{nakazawa14}%
  \BibitemOpen
  \bibfield  {author} {\bibinfo {author} {\bibfnamefont {M.}~\bibnamefont
  {Nakazawa}}, \bibinfo {author} {\bibfnamefont {M.}~\bibnamefont {Yoshida}}, \
  and\ \bibinfo {author} {\bibfnamefont {T.}~\bibnamefont {Hirooka}},\ }\href
  {\doibase 10.1364/OPTICA.1.000015} {\bibfield  {journal} {\bibinfo  {journal}
  {Optica}\ }\textbf {\bibinfo {volume} {1}},\ \bibinfo {pages} {15} (\bibinfo
  {year} {2014})}\BibitemShut {NoStop}%
\bibitem [{\citenamefont {Titulaer}\ and\ \citenamefont
  {Glauber}(1966)}]{tit66}%
  \BibitemOpen
  \bibfield  {author} {\bibinfo {author} {\bibfnamefont {U.}~\bibnamefont
  {Titulaer}}\ and\ \bibinfo {author} {\bibfnamefont {R.}~\bibnamefont
  {Glauber}},\ }\href {http://prola.aps.org/abstract/PR/v145/i4/p1041\_1}
  {\bibfield  {journal} {\bibinfo  {journal} {Phys. Rev.}\ }\textbf {\bibinfo
  {volume} {145}} (\bibinfo {year} {1966})}\BibitemShut {NoStop}%
\bibitem [{\citenamefont {McGuinness}\ \emph {et~al.}(2010)\citenamefont
  {McGuinness}, \citenamefont {Raymer}, \citenamefont {McKinstrie},\ and\
  \citenamefont {Radic}}]{mcg10a}%
  \BibitemOpen
  \bibfield  {author} {\bibinfo {author} {\bibfnamefont {H.~J.}\ \bibnamefont
  {McGuinness}}, \bibinfo {author} {\bibfnamefont {M.~G.}\ \bibnamefont
  {Raymer}}, \bibinfo {author} {\bibfnamefont {C.~J.}\ \bibnamefont
  {McKinstrie}}, \ and\ \bibinfo {author} {\bibfnamefont {S.}~\bibnamefont
  {Radic}},\ }\href {\doibase 10.1103/PhysRevLett.105.093604} {\bibfield
  {journal} {\bibinfo  {journal} {Phys. Rev. Lett.}\ }\textbf {\bibinfo
  {volume} {105}},\ \bibinfo {pages} {093604} (\bibinfo {year}
  {2010})}\BibitemShut {NoStop}%
\bibitem [{\citenamefont {Eckstein}\ \emph {et~al.}(2011)\citenamefont
  {Eckstein}, \citenamefont {Brecht},\ and\ \citenamefont
  {Silberhorn}}]{eck11}%
  \BibitemOpen
  \bibfield  {author} {\bibinfo {author} {\bibfnamefont {A.}~\bibnamefont
  {Eckstein}}, \bibinfo {author} {\bibfnamefont {B.}~\bibnamefont {Brecht}}, \
  and\ \bibinfo {author} {\bibfnamefont {C.}~\bibnamefont {Silberhorn}},\
  }\href@noop {} {\bibfield  {journal} {\bibinfo  {journal} {Opt. Express}\
  }\textbf {\bibinfo {volume} {19}},\ \bibinfo {pages} {13770} (\bibinfo {year}
  {2011})}\BibitemShut {NoStop}%
\bibitem [{\citenamefont {Walmsley}\ and\ \citenamefont {Wong}(1996)}]{wal96}%
  \BibitemOpen
  \bibfield  {author} {\bibinfo {author} {\bibfnamefont {I.~A.}\ \bibnamefont
  {Walmsley}}\ and\ \bibinfo {author} {\bibfnamefont {V.}~\bibnamefont
  {Wong}},\ }\href {\doibase 10.1364/JOSAB.13.002453} {\bibfield  {journal}
  {\bibinfo  {journal} {J. Opt. Soc. Am. B}\ }\textbf {\bibinfo {volume}
  {13}},\ \bibinfo {pages} {2453} (\bibinfo {year} {1996})}\BibitemShut
  {NoStop}%
\bibitem [{\citenamefont {Reddy}\ \emph {et~al.}(2014)\citenamefont {Reddy},
  \citenamefont {Raymer},\ and\ \citenamefont {McKinstrie}}]{red14}%
  \BibitemOpen
  \bibfield  {author} {\bibinfo {author} {\bibfnamefont {D.~V.}\ \bibnamefont
  {Reddy}}, \bibinfo {author} {\bibfnamefont {M.~G.}\ \bibnamefont {Raymer}}, \
  and\ \bibinfo {author} {\bibfnamefont {C.~J.}\ \bibnamefont {McKinstrie}},\
  }\href {\doibase 10.1364/OL.39.002924} {\bibfield  {journal} {\bibinfo
  {journal} {Opt. Lett.}\ }\textbf {\bibinfo {volume} {39}},\ \bibinfo {pages}
  {2924} (\bibinfo {year} {2014})}\BibitemShut {NoStop}%
\bibitem [{\citenamefont {McKinstrie}\ \emph {et~al.}(2012)\citenamefont
  {McKinstrie}, \citenamefont {Mejling}, \citenamefont {Raymer},\ and\
  \citenamefont {Rottwitt}}]{col12}%
  \BibitemOpen
  \bibfield  {author} {\bibinfo {author} {\bibfnamefont {C.~J.}\ \bibnamefont
  {McKinstrie}}, \bibinfo {author} {\bibfnamefont {L.}~\bibnamefont {Mejling}},
  \bibinfo {author} {\bibfnamefont {M.~G.}\ \bibnamefont {Raymer}}, \ and\
  \bibinfo {author} {\bibfnamefont {K.}~\bibnamefont {Rottwitt}},\ }\href
  {\doibase 10.1103/PhysRevA.85.053829} {\bibfield  {journal} {\bibinfo
  {journal} {Phys. Rev. A}\ }\textbf {\bibinfo {volume} {85}},\ \bibinfo
  {pages} {053829} (\bibinfo {year} {2012})}\BibitemShut {NoStop}%
\bibitem [{\citenamefont {Mejling}\ \emph
  {et~al.}(2012{\natexlab{a}})\citenamefont {Mejling}, \citenamefont
  {McKinstrie}, \citenamefont {Raymer},\ and\ \citenamefont
  {Rottwitt}}]{mej12}%
  \BibitemOpen
  \bibfield  {author} {\bibinfo {author} {\bibfnamefont {L.}~\bibnamefont
  {Mejling}}, \bibinfo {author} {\bibfnamefont {C.~J.}\ \bibnamefont
  {McKinstrie}}, \bibinfo {author} {\bibfnamefont {M.~G.}\ \bibnamefont
  {Raymer}}, \ and\ \bibinfo {author} {\bibfnamefont {K.}~\bibnamefont
  {Rottwitt}},\ }\href
  {http://www.opticsinfobase.org/oe/fulltext.cfm?uri=oe-20-8-8367} {\bibfield
  {journal} {\bibinfo  {journal} {Opt. Express}\ }\textbf {\bibinfo {volume}
  {20}},\ \bibinfo {pages} {8367} (\bibinfo {year}
  {2012}{\natexlab{a}})}\BibitemShut {NoStop}%
\bibitem [{\citenamefont {Reddy}\ \emph {et~al.}(2013)\citenamefont {Reddy},
  \citenamefont {Raymer}, \citenamefont {McKinstrie}, \citenamefont {Mejling},\
  and\ \citenamefont {Rottwitt}}]{red13}%
  \BibitemOpen
  \bibfield  {author} {\bibinfo {author} {\bibfnamefont {D.~V.}\ \bibnamefont
  {Reddy}}, \bibinfo {author} {\bibfnamefont {M.~G.}\ \bibnamefont {Raymer}},
  \bibinfo {author} {\bibfnamefont {C.~J.}\ \bibnamefont {McKinstrie}},
  \bibinfo {author} {\bibfnamefont {L.}~\bibnamefont {Mejling}}, \ and\
  \bibinfo {author} {\bibfnamefont {K.}~\bibnamefont {Rottwitt}},\ }\href
  {\doibase 10.1364/OE.21.013840} {\bibfield  {journal} {\bibinfo  {journal}
  {Opt. Express}\ }\textbf {\bibinfo {volume} {21}},\ \bibinfo {pages} {13840}
  (\bibinfo {year} {2013})}\BibitemShut {NoStop}%
\bibitem [{\citenamefont {Kowligy}\ \emph {et~al.}(2014)\citenamefont
  {Kowligy}, \citenamefont {Manurkar}, \citenamefont {Corzo}, \citenamefont
  {Velev}, \citenamefont {Silver}, \citenamefont {Scott}, \citenamefont {Yoo},
  \citenamefont {Kumar}, \citenamefont {Kanter},\ and\ \citenamefont
  {Huang}}]{huang14}%
  \BibitemOpen
  \bibfield  {author} {\bibinfo {author} {\bibfnamefont {A.~S.}\ \bibnamefont
  {Kowligy}}, \bibinfo {author} {\bibfnamefont {P.}~\bibnamefont {Manurkar}},
  \bibinfo {author} {\bibfnamefont {N.~V.}\ \bibnamefont {Corzo}}, \bibinfo
  {author} {\bibfnamefont {V.~G.}\ \bibnamefont {Velev}}, \bibinfo {author}
  {\bibfnamefont {M.}~\bibnamefont {Silver}}, \bibinfo {author} {\bibfnamefont
  {R.~P.}\ \bibnamefont {Scott}}, \bibinfo {author} {\bibfnamefont {S.~J.~B.}\
  \bibnamefont {Yoo}}, \bibinfo {author} {\bibfnamefont {P.}~\bibnamefont
  {Kumar}}, \bibinfo {author} {\bibfnamefont {G.~S.}\ \bibnamefont {Kanter}}, \
  and\ \bibinfo {author} {\bibfnamefont {Y.-P.}\ \bibnamefont {Huang}},\ }\href
  {\doibase 10.1364/OE.22.027942} {\bibfield  {journal} {\bibinfo  {journal}
  {Opt. Express}\ }\textbf {\bibinfo {volume} {22}},\ \bibinfo {pages} {27942}
  (\bibinfo {year} {2014})}\BibitemShut {NoStop}%
\bibitem [{\citenamefont {Raymer}\ and\ \citenamefont
  {Srinivasan}(2012)}]{ptd12}%
  \BibitemOpen
  \bibfield  {author} {\bibinfo {author} {\bibfnamefont {M.~G.}\ \bibnamefont
  {Raymer}}\ and\ \bibinfo {author} {\bibfnamefont {K.}~\bibnamefont
  {Srinivasan}},\ }\href {\doibase 10.1063/PT.3.1786} {\bibfield  {journal}
  {\bibinfo  {journal} {Phys. Today}\ }\textbf {\bibinfo {volume} {65}},\
  \bibinfo {pages} {32} (\bibinfo {year} {2012})}\BibitemShut {NoStop}%
\bibitem [{\citenamefont {Brecht}\ \emph {et~al.}(2014)\citenamefont {Brecht},
  \citenamefont {Eckstein}, \citenamefont {Ricken}, \citenamefont {Quiring},
  \citenamefont {Suche}, \citenamefont {Sansoni},\ and\ \citenamefont
  {Silberhorn}}]{silb14}%
  \BibitemOpen
  \bibfield  {author} {\bibinfo {author} {\bibfnamefont {B.}~\bibnamefont
  {Brecht}}, \bibinfo {author} {\bibfnamefont {A.}~\bibnamefont {Eckstein}},
  \bibinfo {author} {\bibfnamefont {R.}~\bibnamefont {Ricken}}, \bibinfo
  {author} {\bibfnamefont {V.}~\bibnamefont {Quiring}}, \bibinfo {author}
  {\bibfnamefont {H.}~\bibnamefont {Suche}}, \bibinfo {author} {\bibfnamefont
  {L.}~\bibnamefont {Sansoni}}, \ and\ \bibinfo {author} {\bibfnamefont
  {C.}~\bibnamefont {Silberhorn}},\ }\href {\doibase
  10.1103/PhysRevA.90.030302} {\bibfield  {journal} {\bibinfo  {journal} {Phys.
  Rev. A}\ }\textbf {\bibinfo {volume} {90}},\ \bibinfo {pages} {030302}
  (\bibinfo {year} {2014})}\BibitemShut {NoStop}%
\bibitem [{\citenamefont {Ramsey}(1950)}]{ramsey50}%
  \BibitemOpen
  \bibfield  {author} {\bibinfo {author} {\bibfnamefont {N.~F.}\ \bibnamefont
  {Ramsey}},\ }\href {\doibase 10.1103/PhysRev.78.695} {\bibfield  {journal}
  {\bibinfo  {journal} {Phys. Rev.}\ }\textbf {\bibinfo {volume} {78}},\
  \bibinfo {pages} {695} (\bibinfo {year} {1950})}\BibitemShut {NoStop}%
\bibitem [{\citenamefont {Mossberg}\ \emph {et~al.}(1979)\citenamefont
  {Mossberg}, \citenamefont {Kachru}, \citenamefont {Hartmann},\ and\
  \citenamefont {Flusberg}}]{mos79}%
  \BibitemOpen
  \bibfield  {author} {\bibinfo {author} {\bibfnamefont {T.~W.}\ \bibnamefont
  {Mossberg}}, \bibinfo {author} {\bibfnamefont {R.}~\bibnamefont {Kachru}},
  \bibinfo {author} {\bibfnamefont {S.~R.}\ \bibnamefont {Hartmann}}, \ and\
  \bibinfo {author} {\bibfnamefont {A.~M.}\ \bibnamefont {Flusberg}},\ }\href
  {\doibase 10.1103/PhysRevA.20.1976} {\bibfield  {journal} {\bibinfo
  {journal} {Phys. Rev. A}\ }\textbf {\bibinfo {volume} {20}},\ \bibinfo
  {pages} {1976} (\bibinfo {year} {1979})}\BibitemShut {NoStop}%
\bibitem [{\citenamefont {Kasevich}\ \emph {et~al.}(1989)\citenamefont
  {Kasevich}, \citenamefont {Riis}, \citenamefont {Chu},\ and\ \citenamefont
  {DeVoe}}]{chu89}%
  \BibitemOpen
  \bibfield  {author} {\bibinfo {author} {\bibfnamefont {M.~A.}\ \bibnamefont
  {Kasevich}}, \bibinfo {author} {\bibfnamefont {E.}~\bibnamefont {Riis}},
  \bibinfo {author} {\bibfnamefont {S.}~\bibnamefont {Chu}}, \ and\ \bibinfo
  {author} {\bibfnamefont {R.~G.}\ \bibnamefont {DeVoe}},\ }\href {\doibase
  10.1103/PhysRevLett.63.612} {\bibfield  {journal} {\bibinfo  {journal} {Phys.
  Rev. Lett.}\ }\textbf {\bibinfo {volume} {63}},\ \bibinfo {pages} {612}
  (\bibinfo {year} {1989})}\BibitemShut {NoStop}%
\bibitem [{\citenamefont {Clemmen}\ \emph {et~al.}(2014)\citenamefont
  {Clemmen}, \citenamefont {Farsi}, \citenamefont {Ramelow},\ and\
  \citenamefont {Gaeta}}]{gaeta14}%
  \BibitemOpen
  \bibfield  {author} {\bibinfo {author} {\bibfnamefont {S.}~\bibnamefont
  {Clemmen}}, \bibinfo {author} {\bibfnamefont {A.}~\bibnamefont {Farsi}},
  \bibinfo {author} {\bibfnamefont {S.}~\bibnamefont {Ramelow}}, \ and\
  \bibinfo {author} {\bibfnamefont {A.~L.}\ \bibnamefont {Gaeta}},\ }in\ \href
  {\doibase 10.1364/CLEO_QELS.2014.FTh5A.2} {\emph {\bibinfo {booktitle} {CLEO:
  2014 Postdeadline Paper Digest{, paper: FTh5A.2}}}}\ (\bibinfo  {publisher}
  {Optical Society of America},\ \bibinfo {year} {2014})\BibitemShut {NoStop}%
\bibitem [{\citenamefont {Huang}\ and\ \citenamefont {Kumar}(1992)}]{huang92}%
  \BibitemOpen
  \bibfield  {author} {\bibinfo {author} {\bibfnamefont {J.}~\bibnamefont
  {Huang}}\ and\ \bibinfo {author} {\bibfnamefont {P.}~\bibnamefont {Kumar}},\
  }\href {http://link.aps.org/doi/10.1103/PhysRevLett.68.2153} {\bibfield
  {journal} {\bibinfo  {journal} {Phys. Rev. Lett.}\ }\textbf {\bibinfo
  {volume} {68}},\ \bibinfo {pages} {2153} (\bibinfo {year}
  {1992})}\BibitemShut {NoStop}%
\bibitem [{\citenamefont {Vandevender}\ and\ \citenamefont
  {Kwiat}(2004)}]{vand04}%
  \BibitemOpen
  \bibfield  {author} {\bibinfo {author} {\bibfnamefont {A.~P.}\ \bibnamefont
  {Vandevender}}\ and\ \bibinfo {author} {\bibfnamefont {P.~G.}\ \bibnamefont
  {Kwiat}},\ }\href {\doibase 10.1080/09500340410001670884} {\bibfield
  {journal} {\bibinfo  {journal} {J. Mod. Opt.}\ }\textbf {\bibinfo {volume}
  {51}},\ \bibinfo {pages} {1433} (\bibinfo {year} {2004})}\BibitemShut
  {NoStop}%
\bibitem [{\citenamefont {Albota}\ and\ \citenamefont {Wong}(2004)}]{albota04}%
  \BibitemOpen
  \bibfield  {author} {\bibinfo {author} {\bibfnamefont {M.~A.}\ \bibnamefont
  {Albota}}\ and\ \bibinfo {author} {\bibfnamefont {F.~N.~C.}\ \bibnamefont
  {Wong}},\ }\href {http://www.ncbi.nlm.nih.gov/pubmed/15259709} {\bibfield
  {journal} {\bibinfo  {journal} {Opt. Lett.}\ }\textbf {\bibinfo {volume}
  {29}},\ \bibinfo {pages} {1449} (\bibinfo {year} {2004})}\BibitemShut
  {NoStop}%
\bibitem [{\citenamefont {Roussev}\ \emph {et~al.}(2004)\citenamefont
  {Roussev}, \citenamefont {Langrock}, \citenamefont {Kurz},\ and\
  \citenamefont {Fejer}}]{rous04}%
  \BibitemOpen
  \bibfield  {author} {\bibinfo {author} {\bibfnamefont {R.~V.}\ \bibnamefont
  {Roussev}}, \bibinfo {author} {\bibfnamefont {C.}~\bibnamefont {Langrock}},
  \bibinfo {author} {\bibfnamefont {J.~R.}\ \bibnamefont {Kurz}}, \ and\
  \bibinfo {author} {\bibfnamefont {M.~M.}\ \bibnamefont {Fejer}},\ }\href
  {http://www.ncbi.nlm.nih.gov/pubmed/15259732} {\bibfield  {journal} {\bibinfo
   {journal} {Opt. Lett.}\ }\textbf {\bibinfo {volume} {29}},\ \bibinfo {pages}
  {1518} (\bibinfo {year} {2004})}\BibitemShut {NoStop}%
\bibitem [{\citenamefont {Rakher}\ \emph {et~al.}(2010)\citenamefont {Rakher},
  \citenamefont {Ma}, \citenamefont {Slattery}, \citenamefont {Tang},\ and\
  \citenamefont {Srinivasan}}]{rakher10}%
  \BibitemOpen
  \bibfield  {author} {\bibinfo {author} {\bibfnamefont {M.}~\bibnamefont
  {Rakher}}, \bibinfo {author} {\bibfnamefont {L.}~\bibnamefont {Ma}}, \bibinfo
  {author} {\bibfnamefont {O.}~\bibnamefont {Slattery}}, \bibinfo {author}
  {\bibfnamefont {X.}~\bibnamefont {Tang}}, \ and\ \bibinfo {author}
  {\bibfnamefont {K.}~\bibnamefont {Srinivasan}},\ }\href
  {http://dx.doi.org/10.1038/nphoton.2010.221
  http://www.nature.com/nphoton/journal/v4/n11/abs/nphoton.2010.221.html}
  {\bibfield  {journal} {\bibinfo  {journal} {Nature Photon.}\ }\textbf
  {\bibinfo {volume} {4}},\ \bibinfo {pages} {786} (\bibinfo {year}
  {2010})}\BibitemShut {NoStop}%
\bibitem [{\citenamefont {Clark}\ \emph {et~al.}(2013)\citenamefont {Clark},
  \citenamefont {Shahnia}, \citenamefont {Collins}, \citenamefont {Xiong},\
  and\ \citenamefont {Eggleton}}]{cla13}%
  \BibitemOpen
  \bibfield  {author} {\bibinfo {author} {\bibfnamefont {A.~S.}\ \bibnamefont
  {Clark}}, \bibinfo {author} {\bibfnamefont {S.}~\bibnamefont {Shahnia}},
  \bibinfo {author} {\bibfnamefont {M.~J.}\ \bibnamefont {Collins}}, \bibinfo
  {author} {\bibfnamefont {C.}~\bibnamefont {Xiong}}, \ and\ \bibinfo {author}
  {\bibfnamefont {B.~J.}\ \bibnamefont {Eggleton}},\ }\href {\doibase
  10.1364/OL.38.000947} {\bibfield  {journal} {\bibinfo  {journal} {Opt.
  Lett.}\ }\textbf {\bibinfo {volume} {38}},\ \bibinfo {pages} {947} (\bibinfo
  {year} {2013})}\BibitemShut {NoStop}%
\bibitem [{\citenamefont {Myers}\ \emph {et~al.}(1995)\citenamefont {Myers},
  \citenamefont {Eckardt}, \citenamefont {Fejer}, \citenamefont {Byer},
  \citenamefont {Bosenberg},\ and\ \citenamefont {Pierce}}]{myers95}%
  \BibitemOpen
  \bibfield  {author} {\bibinfo {author} {\bibfnamefont {L.~E.}\ \bibnamefont
  {Myers}}, \bibinfo {author} {\bibfnamefont {R.~C.}\ \bibnamefont {Eckardt}},
  \bibinfo {author} {\bibfnamefont {M.~M.}\ \bibnamefont {Fejer}}, \bibinfo
  {author} {\bibfnamefont {R.~L.}\ \bibnamefont {Byer}}, \bibinfo {author}
  {\bibfnamefont {W.~R.}\ \bibnamefont {Bosenberg}}, \ and\ \bibinfo {author}
  {\bibfnamefont {J.~W.}\ \bibnamefont {Pierce}},\ }\href {\doibase
  10.1364/JOSAB.12.002102} {\bibfield  {journal} {\bibinfo  {journal} {J. Opt.
  Soc. Am. B}\ }\textbf {\bibinfo {volume} {12}},\ \bibinfo {pages} {2102}
  (\bibinfo {year} {1995})}\BibitemShut {NoStop}%
\bibitem [{\citenamefont {Mejling}\ \emph
  {et~al.}(2012{\natexlab{b}})\citenamefont {Mejling}, \citenamefont {Cargill},
  \citenamefont {McKinstrie}, \citenamefont {Rottwitt},\ and\ \citenamefont
  {Moore}}]{mej12b}%
  \BibitemOpen
  \bibfield  {author} {\bibinfo {author} {\bibfnamefont {L.}~\bibnamefont
  {Mejling}}, \bibinfo {author} {\bibfnamefont {D.~S.}\ \bibnamefont
  {Cargill}}, \bibinfo {author} {\bibfnamefont {C.~J.}\ \bibnamefont
  {McKinstrie}}, \bibinfo {author} {\bibfnamefont {K.}~\bibnamefont
  {Rottwitt}}, \ and\ \bibinfo {author} {\bibfnamefont {R.~O.}\ \bibnamefont
  {Moore}},\ }\href {\doibase 10.1364/OE.20.027454} {\bibfield  {journal}
  {\bibinfo  {journal} {Opt. Express}\ }\textbf {\bibinfo {volume} {20}},\
  \bibinfo {pages} {27454} (\bibinfo {year} {2012}{\natexlab{b}})}\BibitemShut
  {NoStop}%
\bibitem [{\citenamefont {Raymer}\ \emph {et~al.}(2010)\citenamefont {Raymer},
  \citenamefont {van Enk}, \citenamefont {McKinstrie},\ and\ \citenamefont
  {McGuinness}}]{ray10}%
  \BibitemOpen
  \bibfield  {author} {\bibinfo {author} {\bibfnamefont {M.~G.}\ \bibnamefont
  {Raymer}}, \bibinfo {author} {\bibfnamefont {S.~J.}\ \bibnamefont {van Enk}},
  \bibinfo {author} {\bibfnamefont {C.~J.}\ \bibnamefont {McKinstrie}}, \ and\
  \bibinfo {author} {\bibfnamefont {H.~J.}\ \bibnamefont {McGuinness}},\ }\href
  {\doibase 10.1016/j.optcom.2009.10.057} {\bibfield  {journal} {\bibinfo
  {journal} {Opt. Commun.}\ }\textbf {\bibinfo {volume} {283}},\ \bibinfo
  {pages} {747} (\bibinfo {year} {2010})}\BibitemShut {NoStop}%
\bibitem [{\citenamefont {Burnham}\ and\ \citenamefont
  {Chiao}(1969)}]{Burnham1969}%
  \BibitemOpen
  \bibfield  {author} {\bibinfo {author} {\bibfnamefont {D.~C.}\ \bibnamefont
  {Burnham}}\ and\ \bibinfo {author} {\bibfnamefont {R.~Y.}\ \bibnamefont
  {Chiao}},\ }\href {\doibase 10.1103/PhysRev.188.667} {\bibfield  {journal}
  {\bibinfo  {journal} {Phys. Rev.}\ }\textbf {\bibinfo {volume} {188}},\
  \bibinfo {pages} {667} (\bibinfo {year} {1969})}\BibitemShut {NoStop}%
\bibitem [{\citenamefont {Pfister}\ \emph {et~al.}(2004)\citenamefont
  {Pfister}, \citenamefont {Feng}, \citenamefont {Jennings}, \citenamefont
  {Pooser},\ and\ \citenamefont {Xie}}]{pfister04}%
  \BibitemOpen
  \bibfield  {author} {\bibinfo {author} {\bibfnamefont {O.}~\bibnamefont
  {Pfister}}, \bibinfo {author} {\bibfnamefont {S.}~\bibnamefont {Feng}},
  \bibinfo {author} {\bibfnamefont {G.}~\bibnamefont {Jennings}}, \bibinfo
  {author} {\bibfnamefont {R.}~\bibnamefont {Pooser}}, \ and\ \bibinfo {author}
  {\bibfnamefont {D.}~\bibnamefont {Xie}},\ }\href {\doibase
  10.1103/PhysRevA.70.020302} {\bibfield  {journal} {\bibinfo  {journal} {Phys.
  Rev. A}\ }\textbf {\bibinfo {volume} {70}},\ \bibinfo {pages} {020302}
  (\bibinfo {year} {2004})}\BibitemShut {NoStop}%
\bibitem [{\citenamefont {McKinstrie}\ \emph {et~al.}(2008)\citenamefont
  {McKinstrie}, \citenamefont {van Enk}, \citenamefont {Raymer},\ and\
  \citenamefont {Radic}}]{col08}%
  \BibitemOpen
  \bibfield  {author} {\bibinfo {author} {\bibfnamefont {C.~J.}\ \bibnamefont
  {McKinstrie}}, \bibinfo {author} {\bibfnamefont {S.~J.}\ \bibnamefont {van
  Enk}}, \bibinfo {author} {\bibfnamefont {M.~G.}\ \bibnamefont {Raymer}}, \
  and\ \bibinfo {author} {\bibfnamefont {S.}~\bibnamefont {Radic}},\ }\href
  {\doibase 10.1364/OE.16.002720} {\bibfield  {journal} {\bibinfo  {journal}
  {Opt. Express}\ }\textbf {\bibinfo {volume} {16}},\ \bibinfo {pages} {2720}
  (\bibinfo {year} {2008})}\BibitemShut {NoStop}%
\bibitem [{\citenamefont {{U'Ren}}\ \emph {et~al.}(2005)\citenamefont
  {{U'Ren}}, \citenamefont {Silberhorn}, \citenamefont {Banaszek},
  \citenamefont {Walmsley}, \citenamefont {Erdmann}, \citenamefont {Grice},\
  and\ \citenamefont {Raymer}}]{uren05}%
  \BibitemOpen
  \bibfield  {author} {\bibinfo {author} {\bibfnamefont {A.~B.}\ \bibnamefont
  {{U'Ren}}}, \bibinfo {author} {\bibfnamefont {C.}~\bibnamefont {Silberhorn}},
  \bibinfo {author} {\bibfnamefont {K.}~\bibnamefont {Banaszek}}, \bibinfo
  {author} {\bibfnamefont {I.~A.}\ \bibnamefont {Walmsley}}, \bibinfo {author}
  {\bibfnamefont {R.}~\bibnamefont {Erdmann}}, \bibinfo {author} {\bibfnamefont
  {W.~P.}\ \bibnamefont {Grice}}, \ and\ \bibinfo {author} {\bibfnamefont
  {M.~G.}\ \bibnamefont {Raymer}},\ }\href@noop {} {\bibfield  {journal}
  {\bibinfo  {journal} {Laser Phys.}\ }\textbf {\bibinfo {volume} {15}},\
  \bibinfo {pages} {146} (\bibinfo {year} {2005})}\BibitemShut {NoStop}%
\bibitem [{\citenamefont {S\o{}ndergaard}\ and\ \citenamefont
  {Dridi}(2000)}]{drid00}%
  \BibitemOpen
  \bibfield  {author} {\bibinfo {author} {\bibfnamefont {T.}~\bibnamefont
  {S\o{}ndergaard}}\ and\ \bibinfo {author} {\bibfnamefont {K.~H.}\
  \bibnamefont {Dridi}},\ }\href {\doibase 10.1103/PhysRevB.61.15688}
  {\bibfield  {journal} {\bibinfo  {journal} {Phys. Rev. B}\ }\textbf {\bibinfo
  {volume} {61}},\ \bibinfo {pages} {15688} (\bibinfo {year}
  {2000})}\BibitemShut {NoStop}%
\bibitem [{\citenamefont {Mejling}\ \emph {et~al.}(2014)\citenamefont
  {Mejling}, \citenamefont {Friis}, \citenamefont {Reddy}, \citenamefont
  {Rottwitt}, \citenamefont {Raymer},\ and\ \citenamefont
  {McKinstrie}}]{mej14}%
  \BibitemOpen
  \bibfield  {author} {\bibinfo {author} {\bibfnamefont {L.}~\bibnamefont
  {Mejling}}, \bibinfo {author} {\bibfnamefont {S.~M.~M.}\ \bibnamefont
  {Friis}}, \bibinfo {author} {\bibfnamefont {D.~V.}\ \bibnamefont {Reddy}},
  \bibinfo {author} {\bibfnamefont {K.}~\bibnamefont {Rottwitt}}, \bibinfo
  {author} {\bibfnamefont {M.~G.}\ \bibnamefont {Raymer}}, \ and\ \bibinfo
  {author} {\bibfnamefont {C.~J.}\ \bibnamefont {McKinstrie}},\ }in\ \href
  {http://www.opticsinfobase.org/abstract.cfm?URI=NP-2014-JTu3A.36} {\emph
  {\bibinfo {booktitle} {Advanced Photon.{, paper: JTu3A.36}}}}\ (\bibinfo
  {publisher} {Optical Society of America},\ \bibinfo {year}
  {2014})\BibitemShut {NoStop}%
\bibitem [{\citenamefont {Hayat}\ \emph {et~al.}(2012)\citenamefont {Hayat},
  \citenamefont {Xing}, \citenamefont {Feizpour},\ and\ \citenamefont
  {Steinberg}}]{hayat}%
  \BibitemOpen
  \bibfield  {author} {\bibinfo {author} {\bibfnamefont {A.}~\bibnamefont
  {Hayat}}, \bibinfo {author} {\bibfnamefont {X.}~\bibnamefont {Xing}},
  \bibinfo {author} {\bibfnamefont {A.}~\bibnamefont {Feizpour}}, \ and\
  \bibinfo {author} {\bibfnamefont {A.~M.}\ \bibnamefont {Steinberg}},\ }\href
  {\doibase 10.1364/OE.20.029174} {\bibfield  {journal} {\bibinfo  {journal}
  {Opt. Express}\ }\textbf {\bibinfo {volume} {20}},\ \bibinfo {pages} {29174}
  (\bibinfo {year} {2012})}\BibitemShut {NoStop}%
\bibitem [{\citenamefont {Heritage}\ and\ \citenamefont
  {Weiner}(2007)}]{herit}%
  \BibitemOpen
  \bibfield  {author} {\bibinfo {author} {\bibfnamefont {J.}~\bibnamefont
  {Heritage}}\ and\ \bibinfo {author} {\bibfnamefont {A.}~\bibnamefont
  {Weiner}},\ }\href {\doibase 10.1109/JSTQE.2007.901891} {\bibfield  {journal}
  {\bibinfo  {journal} {IEEE J. Sel. Top. Quantum Electron.}\ }\textbf
  {\bibinfo {volume} {13}},\ \bibinfo {pages} {1351} (\bibinfo {year}
  {2007})}\BibitemShut {NoStop}%
\bibitem [{\citenamefont {Brecht}\ \emph {et~al.}(2011)\citenamefont {Brecht},
  \citenamefont {Eckstein}, \citenamefont {Christ}, \citenamefont {Suche},\
  and\ \citenamefont {Silberhorn}}]{brecht11}%
  \BibitemOpen
  \bibfield  {author} {\bibinfo {author} {\bibfnamefont {B.}~\bibnamefont
  {Brecht}}, \bibinfo {author} {\bibfnamefont {A.}~\bibnamefont {Eckstein}},
  \bibinfo {author} {\bibfnamefont {A.}~\bibnamefont {Christ}}, \bibinfo
  {author} {\bibfnamefont {H.}~\bibnamefont {Suche}}, \ and\ \bibinfo {author}
  {\bibfnamefont {C.}~\bibnamefont {Silberhorn}},\ }\href
  {http://stacks.iop.org/1367-2630/13/i=6/a=065029} {\bibfield  {journal}
  {\bibinfo  {journal} {New J. Phys.}\ }\textbf {\bibinfo {volume} {13}},\
  \bibinfo {pages} {065029} (\bibinfo {year} {2011})}\BibitemShut {NoStop}%
\bibitem [{\citenamefont {Nunn}\ \emph {et~al.}(2013)\citenamefont {Nunn},
  \citenamefont {Wright}, \citenamefont {S\"{o}ller}, \citenamefont {Zhang},
  \citenamefont {Walmsley},\ and\ \citenamefont {Smith}}]{nunn13}%
  \BibitemOpen
  \bibfield  {author} {\bibinfo {author} {\bibfnamefont {J.}~\bibnamefont
  {Nunn}}, \bibinfo {author} {\bibfnamefont {L.~J.}\ \bibnamefont {Wright}},
  \bibinfo {author} {\bibfnamefont {C.}~\bibnamefont {S\"{o}ller}}, \bibinfo
  {author} {\bibfnamefont {L.}~\bibnamefont {Zhang}}, \bibinfo {author}
  {\bibfnamefont {I.~A.}\ \bibnamefont {Walmsley}}, \ and\ \bibinfo {author}
  {\bibfnamefont {B.~J.}\ \bibnamefont {Smith}},\ }\href {\doibase
  10.1364/OE.21.015959} {\bibfield  {journal} {\bibinfo  {journal} {Opt.
  Express}\ }\textbf {\bibinfo {volume} {21}},\ \bibinfo {pages} {15959}
  (\bibinfo {year} {2013})}\BibitemShut {NoStop}%
\bibitem [{\citenamefont {Jensen}\ \emph {et~al.}(2011)\citenamefont {Jensen},
  \citenamefont {Wasilewski}, \citenamefont {Krauter}, \citenamefont
  {Fernholz}, \citenamefont {Nielsen}, \citenamefont {Owari}, \citenamefont
  {Plenio}, \citenamefont {Serafini}, \citenamefont {Wolf},\ and\ \citenamefont
  {Polzik}}]{polzik11}%
  \BibitemOpen
  \bibfield  {author} {\bibinfo {author} {\bibfnamefont {K.}~\bibnamefont
  {Jensen}}, \bibinfo {author} {\bibfnamefont {W.}~\bibnamefont {Wasilewski}},
  \bibinfo {author} {\bibfnamefont {H.}~\bibnamefont {Krauter}}, \bibinfo
  {author} {\bibfnamefont {T.}~\bibnamefont {Fernholz}}, \bibinfo {author}
  {\bibfnamefont {B.~M.}\ \bibnamefont {Nielsen}}, \bibinfo {author}
  {\bibfnamefont {M.}~\bibnamefont {Owari}}, \bibinfo {author} {\bibfnamefont
  {M.~B.}\ \bibnamefont {Plenio}}, \bibinfo {author} {\bibfnamefont
  {A.}~\bibnamefont {Serafini}}, \bibinfo {author} {\bibfnamefont {M.~M.}\
  \bibnamefont {Wolf}}, \ and\ \bibinfo {author} {\bibfnamefont {E.~S.}\
  \bibnamefont {Polzik}},\ }\href@noop {} {\bibfield  {journal} {\bibinfo
  {journal} {Nature Phys.}\ }\textbf {\bibinfo {volume} {7}},\ \bibinfo {pages}
  {13} (\bibinfo {year} {2011})}\BibitemShut {NoStop}%
\bibitem [{\citenamefont {Humphreys}\ \emph {et~al.}(2014)\citenamefont
  {Humphreys}, \citenamefont {Kolthammer}, \citenamefont {Nunn}, \citenamefont
  {Barbieri}, \citenamefont {Datta},\ and\ \citenamefont {Walmsley}}]{humph14}%
  \BibitemOpen
  \bibfield  {author} {\bibinfo {author} {\bibfnamefont {P.~C.}\ \bibnamefont
  {Humphreys}}, \bibinfo {author} {\bibfnamefont {W.~S.}\ \bibnamefont
  {Kolthammer}}, \bibinfo {author} {\bibfnamefont {J.}~\bibnamefont {Nunn}},
  \bibinfo {author} {\bibfnamefont {M.}~\bibnamefont {Barbieri}}, \bibinfo
  {author} {\bibfnamefont {A.}~\bibnamefont {Datta}}, \ and\ \bibinfo {author}
  {\bibfnamefont {I.~A.}\ \bibnamefont {Walmsley}},\ }\href {\doibase
  10.1103/PhysRevLett.113.130502} {\bibfield  {journal} {\bibinfo  {journal}
  {Phys. Rev. Lett.}\ }\textbf {\bibinfo {volume} {113}},\ \bibinfo {pages}
  {130502} (\bibinfo {year} {2014})}\BibitemShut {NoStop}%
\bibitem [{\citenamefont {Humphreys}\ \emph {et~al.}(2013)\citenamefont
  {Humphreys}, \citenamefont {Metcalf}, \citenamefont {Spring}, \citenamefont
  {Moore}, \citenamefont {Jin}, \citenamefont {Barbieri}, \citenamefont
  {Kolthammer},\ and\ \citenamefont {Walmsley}}]{humph13}%
  \BibitemOpen
  \bibfield  {author} {\bibinfo {author} {\bibfnamefont {P.~C.}\ \bibnamefont
  {Humphreys}}, \bibinfo {author} {\bibfnamefont {B.~J.}\ \bibnamefont
  {Metcalf}}, \bibinfo {author} {\bibfnamefont {J.~B.}\ \bibnamefont {Spring}},
  \bibinfo {author} {\bibfnamefont {M.}~\bibnamefont {Moore}}, \bibinfo
  {author} {\bibfnamefont {X.-M.}\ \bibnamefont {Jin}}, \bibinfo {author}
  {\bibfnamefont {M.}~\bibnamefont {Barbieri}}, \bibinfo {author}
  {\bibfnamefont {W.~S.}\ \bibnamefont {Kolthammer}}, \ and\ \bibinfo {author}
  {\bibfnamefont {I.~A.}\ \bibnamefont {Walmsley}},\ }\href {\doibase
  10.1103/PhysRevLett.111.150501} {\bibfield  {journal} {\bibinfo  {journal}
  {Phys. Rev. Lett.}\ }\textbf {\bibinfo {volume} {111}},\ \bibinfo {pages}
  {150501} (\bibinfo {year} {2013})}\BibitemShut {NoStop}%
\bibitem [{\citenamefont {Cundiff}\ and\ \citenamefont
  {Weiner}(2010)}]{cundiff10}%
  \BibitemOpen
  \bibfield  {author} {\bibinfo {author} {\bibfnamefont {S.~T.}\ \bibnamefont
  {Cundiff}}\ and\ \bibinfo {author} {\bibfnamefont {A.~M.}\ \bibnamefont
  {Weiner}},\ }\href {\doibase nphoton.2010.196} {\bibfield  {journal}
  {\bibinfo  {journal} {Nature Photon.}\ }\textbf {\bibinfo {volume} {4}},\
  \bibinfo {pages} {760} (\bibinfo {year} {2010})}\BibitemShut {NoStop}%
\bibitem [{\citenamefont {Weiner}(2011)}]{weiner11}%
  \BibitemOpen
  \bibfield  {author} {\bibinfo {author} {\bibfnamefont {A.~M.}\ \bibnamefont
  {Weiner}},\ }\href {\doibase http://dx.doi.org/10.1016/j.optcom.2011.03.084}
  {\bibfield  {journal} {\bibinfo  {journal} {Optics Commun.}\ }\textbf
  {\bibinfo {volume} {284}},\ \bibinfo {pages} {3669 } (\bibinfo {year}
  {2011})}\BibitemShut {NoStop}%
\end{thebibliography}

%

\end{document}